%% file: paper.tex
\documentclass[conference,a4paper]{IEEEtran}

\usepackage[utf8]{inputenc} 
\usepackage[T1]{fontenc}
\usepackage{url}              
\usepackage{cite}      

\usepackage[cmex10]{amsmath}  
\usepackage{mleftright}    
\mleftright                   

\usepackage{graphicx}      
\usepackage{booktabs}         

\usepackage[caption=false,font=footnotesize]{subfig}

\usepackage{diagbox}
\usepackage{graphicx}
\usepackage{multirow}

\def\v {\mathbf{v}}

\def\G {\mathcal{G}}

\input{math_commands.tex}

\newtheorem{defn}{Definition}

\begin{document}

\title{Learning to Code on Graphs for Topological Interference Management} 

\author{%
\IEEEauthorblockN{Zhiwei Shan\IEEEauthorrefmark{1},
                Xinping Yi\IEEEauthorrefmark{1},
                    Han Yu\IEEEauthorrefmark{2},
                    Chung-Shou Liao\IEEEauthorrefmark{3},
                    and Shi Jin\IEEEauthorrefmark{4}}
\IEEEauthorblockA{\IEEEauthorrefmark{1}%
                    University of Liverpool, Liverpool L69 3GJ, England, 
                    \{zshan, xinping.yi\}@liverpool.ac.uk}
\IEEEauthorblockA{\IEEEauthorrefmark{2}%
                     Chalmers University of Technology, Gothenburg SE41296, Sweden,
                     yuha@chalmers.se}
\IEEEauthorblockA{\IEEEauthorrefmark{3}%
                     National Tsing Hua University, 
                     Hsinchu 30013, Taiwan,
                     csliao@ie.nthu.edu.tw}
\IEEEauthorblockA{\IEEEauthorrefmark{4}%
                     Southeast University, 
                     Nanjing 210096, China,
                     jinshi@seu.edu.cn}
 }

\maketitle

\begin{abstract}

  The state-of-the-art coding schemes for topological interference management (TIM) problems are usually handcrafted for specific families of network topologies, relying critically on experts’ domain knowledge.
  This inevitably restricts the potential wider applications to wireless communication systems, due to the limited generalizability.
  This work makes the first attempt to advocate a novel intelligent
  coding approach to mimic topological interference alignment (IA) via local graph coloring algorithms, leveraging the new advances of graph neural networks (GNNs) and reinforcement learning (RL). 
  The proposed LCG framework is then generalized to discover new IA coding schemes, including one-to-one vector IA and subspace IA.
  The extensive experiments demonstrate the excellent generalizability and transferability of the proposed approach, where the parameterized GNNs trained by small size TIM instances are able to work well on new unseen network topologies with larger size.
\end{abstract}

\section{Introduction}
\label{sec:Introduction}

Topological interference management (TIM) is one of the most promising techniques for wireless networks with much relaxed requirement of channel state information (CSI) at the transmitters. As introduced in \cite{jafar2013topological}, TIM examines the degrees of freedom (DoF) of partially connected one-hop wireless networks with the only available CSI at the transmitters being the network topology. Over the past few years, TIM has received extensive attention, resulting in a growing number of follow-up works, including TIM with alternating topology \cite{sun2013alternating,gherekhloo2013alternating2}, multi-level TIM \cite{geng2013multilevel,geng2021multilevel}, TIM with cooperation \cite{yi2015topological,yi2018topological}, dynamic TIM \cite{yi2019opportunistic,liang2022topological}, and many others (e.g., \cite{multi-antenna,maleki2013optimality,yi2018tdma,doumiati2019framework,davoodi2018network,el2017topological,yang2017topological,aquilina2016degrees,shi2016low,gao2014topological,naderializadeh2014interference,mutangana2020topological}). 
It is worth noting that interference alignment, as a simple yet elegant linear coding technique, has been proven to have theoretical potential to improve over the conventional orthogonal access approaches, such as TDMA, frequency reuse, and CDMA \cite{jafar2013topological}. 

However, state-of-the-art coding techniques for TIM inspect specific network topologies individually and design coding schemes in a handcrafted manner. This relies critically on experts' domain knowledge, which may be time-consuming, ungeneralizable, and unscalable, restricting the wide applications to wireless system designs. As machine learning has been increasingly involved in wireless system design, one may wonder if learning can be leveraged to design and discover new coding techniques. Some initial attempts have been made in the literature (e.g., \cite{kim2018communication,kim2020deepcode,jiang2019turbo,he2020model,9517735,chahine2021deepic,zhang2020factor,satorras2021neural}), where deep neural networks have been employed to design new neural decoders. Nevertheless, it is still unclear whether or not machine learning could be applied to coding on graphs, where the graph structures impose challenges on learning to code.

\begin{figure*}[h]
	\begin{center}
  \includegraphics[width=0.4\textwidth]{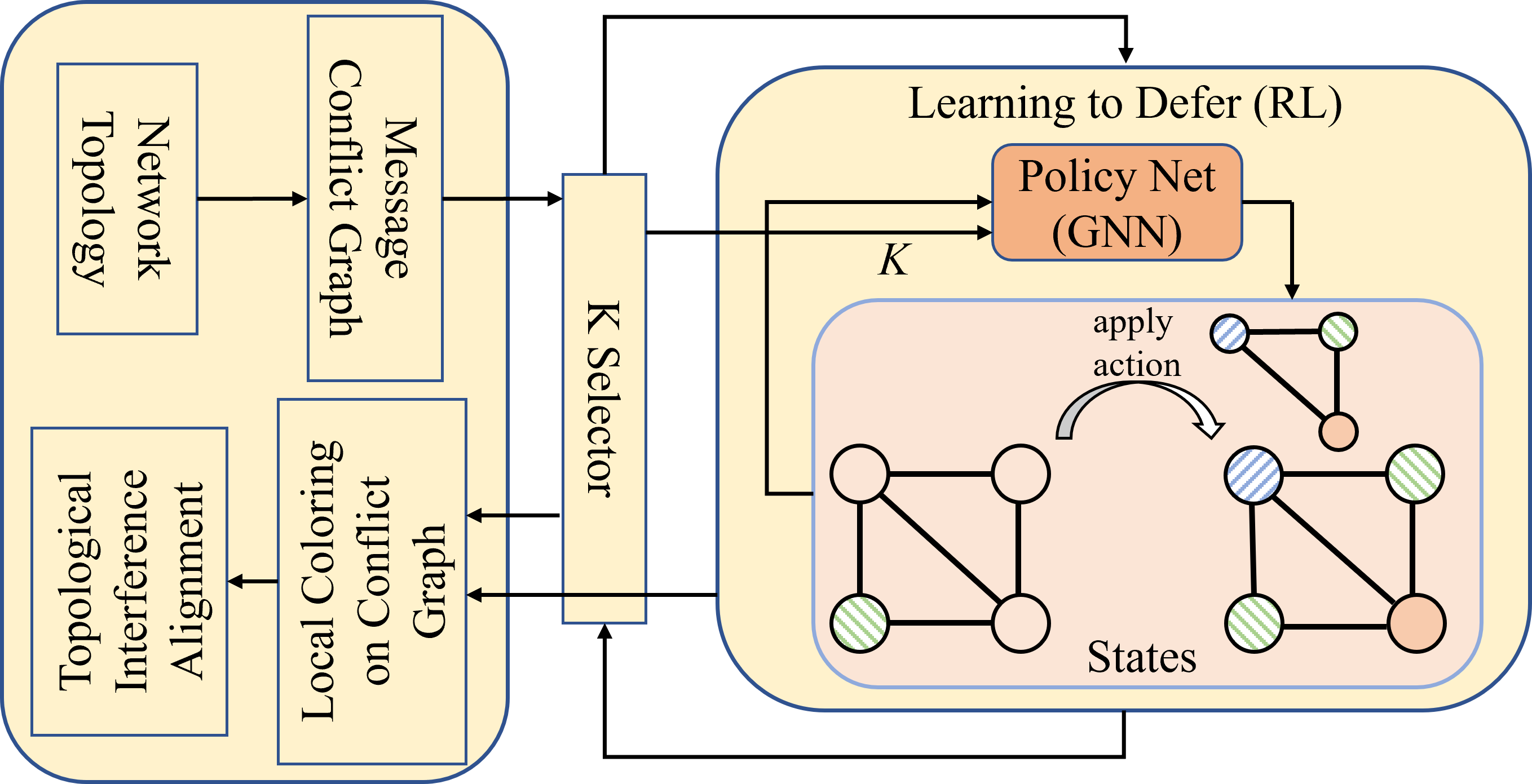}
		\includegraphics[width=0.52\textwidth]{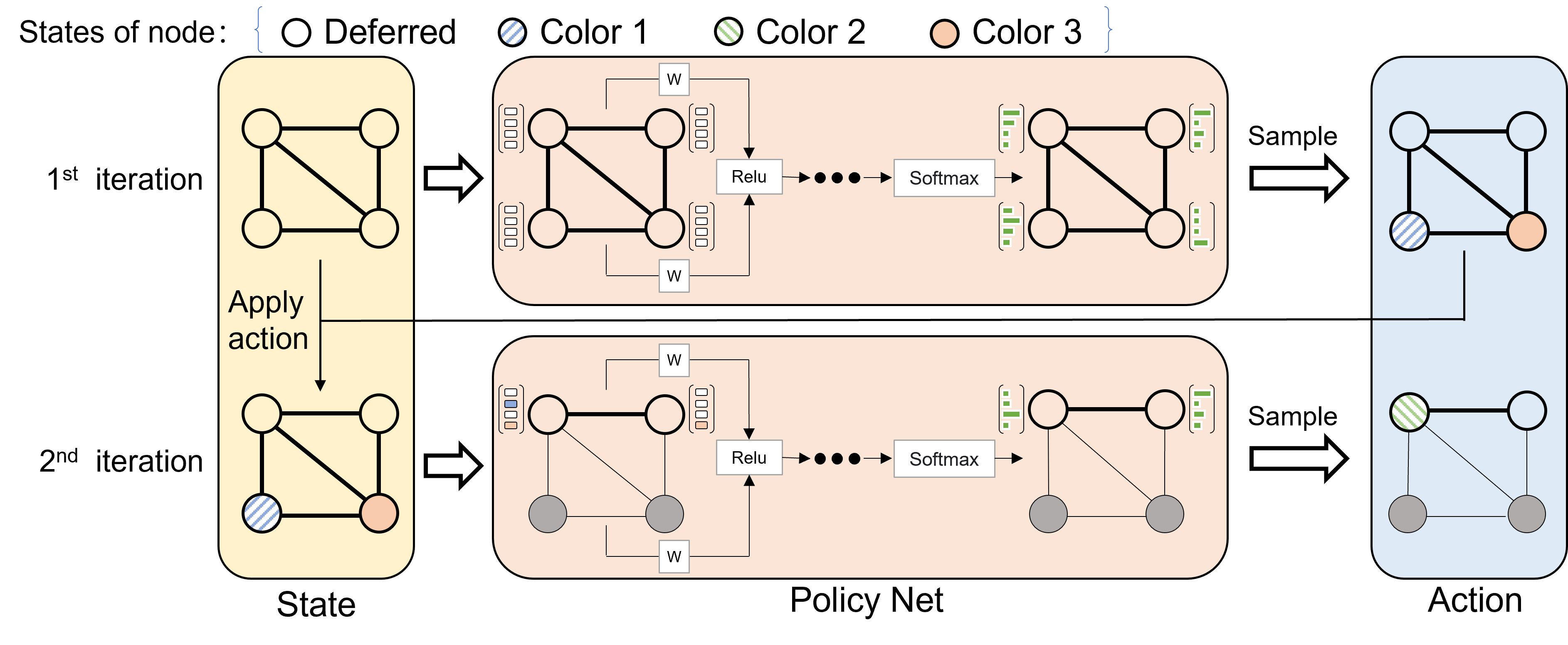}
	\end{center}
	\caption{(a) The framework of the proposed Learning to Code on Graphs (LCG), where vector assignment for one-to-one IA is done by local coloring on conflict graph using a learning-to-defer approach. (b) The iterative procedure of learning to defer for local coloring.}
	\label{1st framework}
\end{figure*}

In this paper, we make a first attempt to push forward this line of research, taking the TIM problem as an example to propose a novel intelligent coding framework for learning-to-code on graphs (LCG). The proposed LCG framework takes interference alignment (IA) as the template, 
translating beamforming vector design of IA into dedicated vector generation followed by vector assignment according to IA conditions.
The key ingredient of LCG is an intelligent combinatorial optimization algorithm, which leverages reinforcement learning (RL) for vector assignment and Graph Neural Network (GNN) for graph representation learning on conflict graphs.
As shown in Figure \ref{1st framework}, for TIM instances, the directed message conflict graphs are first constructed from network topologies, followed by a learning-to-defer approach to assigning vectors to the conflict graph in an iterative manner with state transition until certain IA conditions are satisfied. Such a vector assignment strategy will be translated to topological IA for beamforming.

To be specific, we consider four types of IA: one-to-one scalar IA (OSIA), one-to-one vector IA (OVIA), subspace scalar IA (SSIA), and subspace vector IA (SVIA). 
For one-to-one IA, we relate OSIA and OVIA to local graph coloring \cite{shanmugam2013local} and fractional local graph coloring, respectively (see Sec. \ref{Local Coloring}, \ref{Fractional Local Coloring}), in such a way that LCG with a learning-to-defer approach can mimic graph coloring algorithms to learn a valid local coloring strategy for assigning beamforming vectors (cf. colors) to messages, achieving one-to-one IA. The beamforming vectors can be generated by maximum distance separable (MDS) codes.
 
However, when it comes to subspace IA, specialized vector generation is required as MDS codes with linear independence is insufficient to satisfy subspace IA conditions, and vector assignment via graph coloring is inadequate to specify the overlap of subspace for partial IA.
To overcome these issues, we generate beamforming vectors in an implicit way to distinguish different subspace and propose a new approach called Matrix Rank Reduction for vector assignment, which is specifically tailored to meet the conditions of subspace IA (see Sec. \ref{Subspace Interference Alignment via Matrix Rank Reduction}). In doing so, the learn-to-defer approach for local coloring can be effectively extended to
learn a valid strategy for directly assigning beamforming matrices to messages, achieving the desired subspace alignment.

To evaluate the effectiveness of the proposed LCG framework, we conduct extensive experiments on various TIM instances with random network topologies.
Experimental results show that the proposed graph coloring method is more time-efficient and effective than traditional methods including smallest-last greedy method \cite{matula1983smallest} with interchange \cite{deo2006interchange} and TabuCol \cite{hertz1987using} on coloring problems. 
We also tested our models on various types of IA tasks and the results confirmed that our proposed method can be used to automatically discover those IA coding schemes, and theoretically superior IA methods can indeed achieve higher DoF. 
We finally experimentally demonstrate the strong transferability of our model. It is evidenced that the trained model with small-size random graphs can be generalized to unseen graph types with larger sizes without loss of performance. 

The rest of this paper is organized as follows. Section II provides the statement of the TIM problem and describes four types of IA coding schemes. The translations to vector assignment via local coloring for one-to-one IA and via matrix rank reduction for subspace IA are considered in Section III. Section IV is dedicated to our proposed LCG framework for local coloring and matrix rank reduction, followed by detailed experimental setups and evalution results in Section V. Finally we conclude the paper in Section VI.

\section{Problem Statement}
\subsection{Topological Interference Management}

The TIM problem considers a partially connected interference network that has $M$ sources, labeled as $S_1,S_2,\dots,S_M$, and $N$ destinations, labeled as $D_1,D_2,\dots,D_N$, with each equipped one single antenna. The topology matrix $\mT$ is a $N \times M$ matrix with elements $t_{ji} = 1$ if there exists a non-zero channel from Source $i$ to Destination $j$ and $t_{ji} = 0$ otherwise. At time instant $t$, 
Destination $D_j$ receives signal:
\begin{align}
Y_j(t) = \sum_{i=1}^Mt_{ji}h_{ji}X_i(t)+Z_j(t),
\end{align}
where $X_i(t)$ is the symbol transmitted by Source $S_i$. All transmitted signals are subject to a power constraint $P$. $h_{ij}$ is the constant channel coefficient between Source $S_i$ and Destination $D_j$, $Z_j(t)$ is the additive white Gaussian noise (AWGN) at Destination $D_j$, and $Y_j(t)$ is the symbol received by Destination $D_j$. All symbols are complex-valued. The topology matrix $\mT$ is known by all sources and destinations. 
Throughout this paper, we focus on the multiple unicast setting, where each source is paired with a unique destination with one desired message delivered.

\begin{figure}[htbp]
   \centering
   \vspace{-20pt}
   \subfloat[Network topology graph.]{%
     \label{5nodes_sample_topo}
     \includegraphics[width=3.1cm]{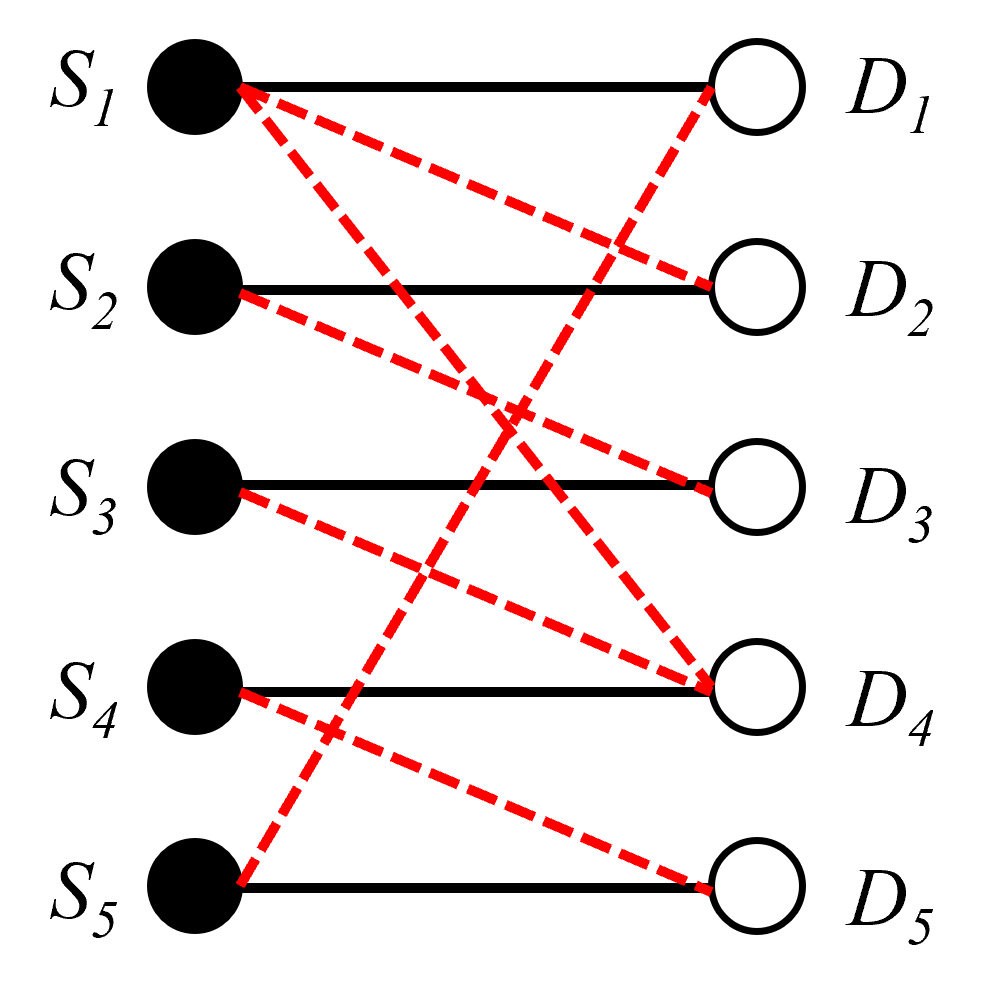}}
   \hfil
   \subfloat[Message conflict graph.]{%
     \label{5nodes_sample_conf}
     \includegraphics[width=3.5cm]{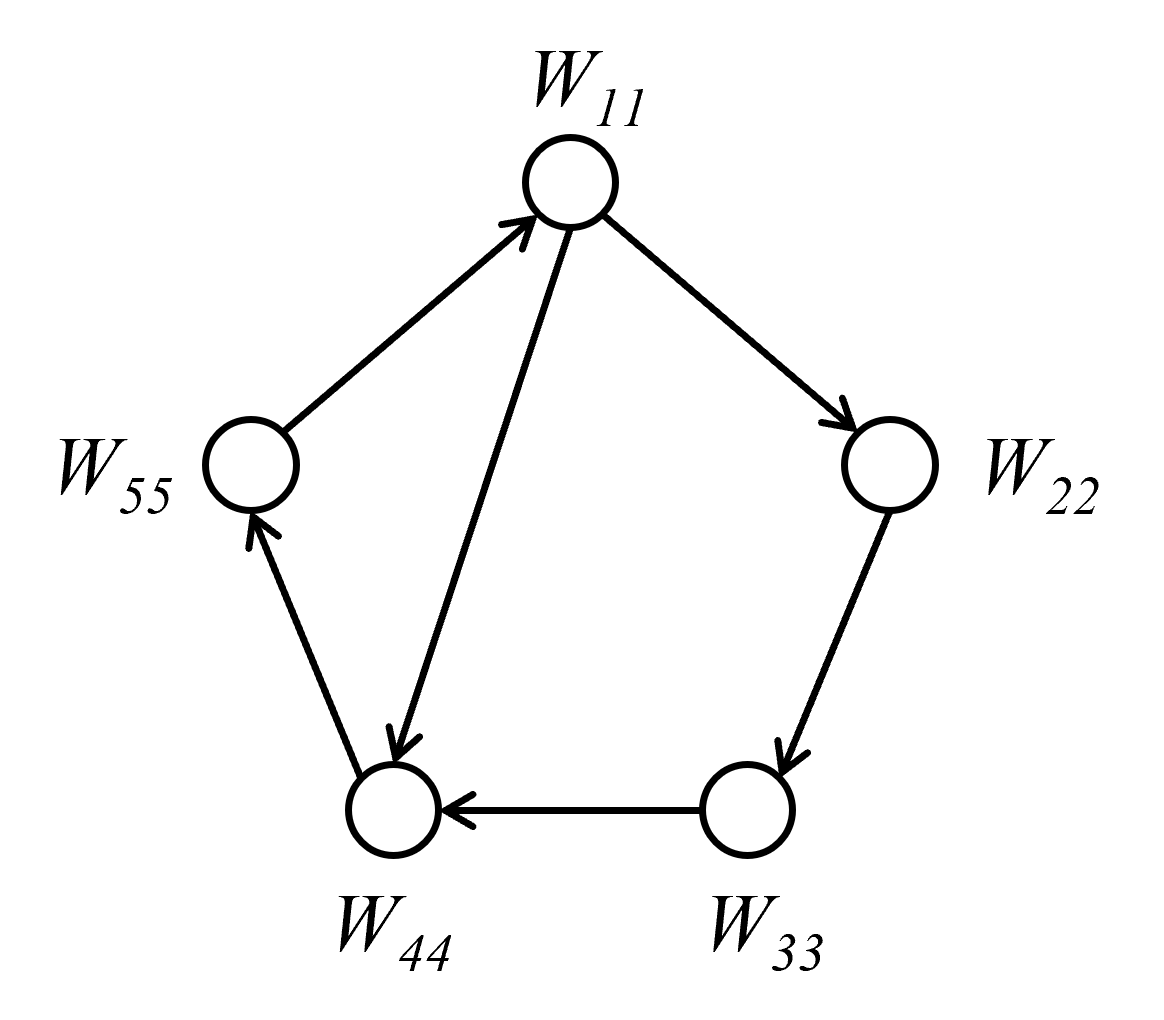}}  
   \caption{(a) A 5-node TIM instance topology graph, where the black edges indicate paired sources and destinations with desired messages and the red dotted edges are interfering signals, and (b) the corresponding message conflict graph with desired messages (i.e., source-destination pairs) being vertices and the directed edges indicate interference from sources to destinations.}
   \label{5_nodes_sample}
\end{figure}

We introduce two graph definitions for TIM representation.
\begin{defn}
	Given the TIM problem with $M$ sources and $N$ destinations, topology matrix $\mT$, and message set $\gM$, define the following two graphs:
	\begin{itemize}
		\item [1)] 
		\textbf{Network Topology Graph}: An undirected bipartite graph
		with sources on one side, destinations on the other, and
		an edge between $S_i$ and $D_j$ whenever $t_{ji} = 1$.     
		\item [2)]
		\textbf{Message Conflict Graph}: An directed graph where
		each message $W_{ji} \in \gM$ is a vertex, and a directed edge $(W_{ji}, W_{j'i'})$ exists if and only if $D_{j'}$ is interfered by $S_i$, i.e., $t_{ij'} = 1$, where $\gM$ is the set of desired messages.
	\end{itemize}
\end{defn}
An example is shown in Figure \ref{5_nodes_sample}. Note that a connection between a source $S$ and a destination $D$ could be one of two cases: demanded link and interfering link. The link is demanded if there exists a desired message from $S$ to $D$ (black solid lines in Figure \ref{5nodes_sample_topo}) and interfering otherwise (red dotted lines in Figure \ref{5nodes_sample_topo}).
A similar undirected version of message conflict graph was defined in \cite{yi2018tdma}, which ignores some information of conflicting source.

One of the major objectives of the TIM problem is to maximize the symmetric degrees-of-freedom (DoFs), $d_{\mathrm{sym}}$, which refer to the minimal pre-log value of the achievable rate over all demanded messages $\gM$. For rigorous definitions, please refer to \cite{jafar2013topological} for details.

\subsection{Interference Alignment Perspective}\label{Interference Alignment Perspective}
The majority of TIM solutions are linear coding schemes, which can be roughly divided into two categories: scalar and vector linear coding.
The scalar coding scheme transmits one symbol for each message over $K$ channel uses, yielding DoF of $d=\frac{1}{K}$. In contrast, the vector coding scheme divides the message into $b$ independent scalar streams, with each stream carrying one symbol. These symbols are then transmitted along the corresponding column vectors of the pre-coding matrix $\mathbf{V} \in \mathbb{C}^{K \times b}$, also known as the beamforming vectors, over $K$ channel uses, yielding DoF of $d=\frac{b}{K}$.  
The vector coding enables simultaneous transmission of multiple symbols, making full use of the spatial-time dimension of the channel, thus improving spectral efficiency of the communication system.

\begin{figure*}
   \centering
   \vspace{-20pt}
   \subfloat[Network topology graph.]{%
     \label{5nodes_sample_topo_11vector}
     \includegraphics[width=3.1cm]{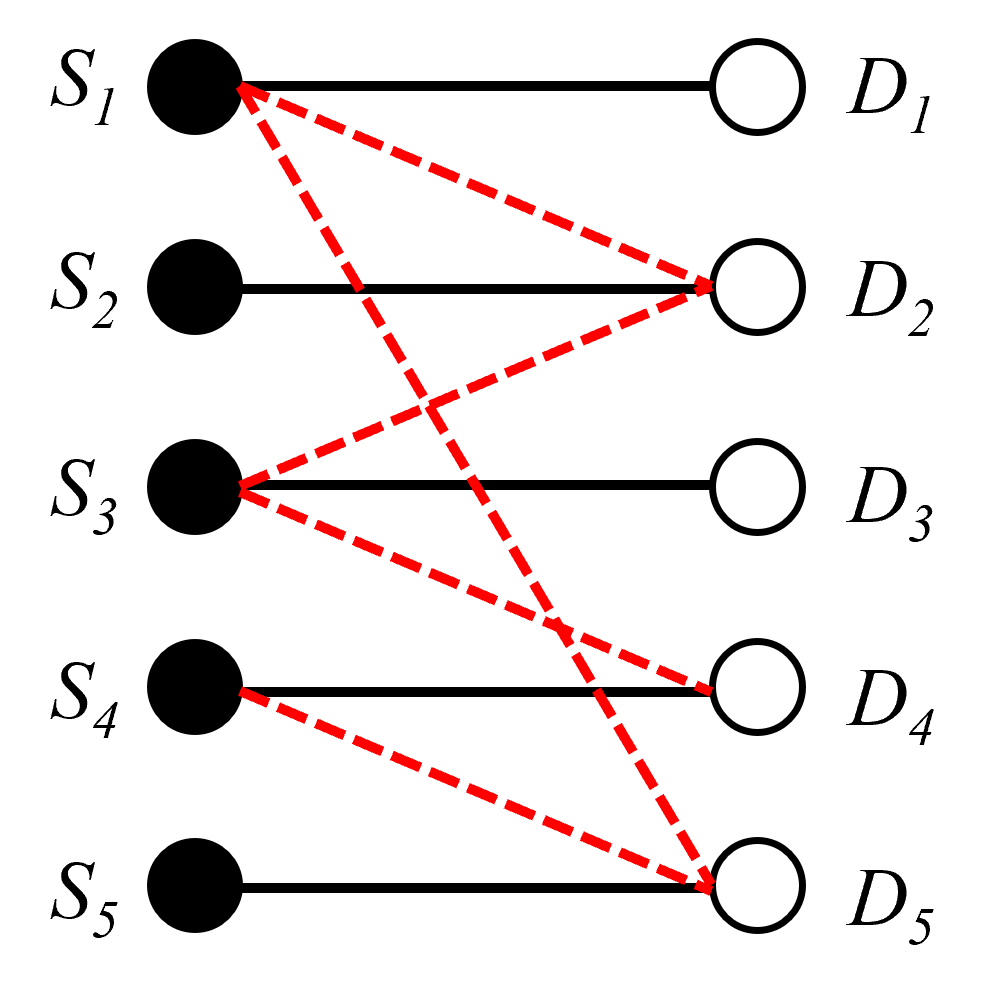}}
   \hfil
   \subfloat[Message conflict graph.]{%
     \label{5nodes_sample_conf_11vector}
     \includegraphics[width=3.5cm]{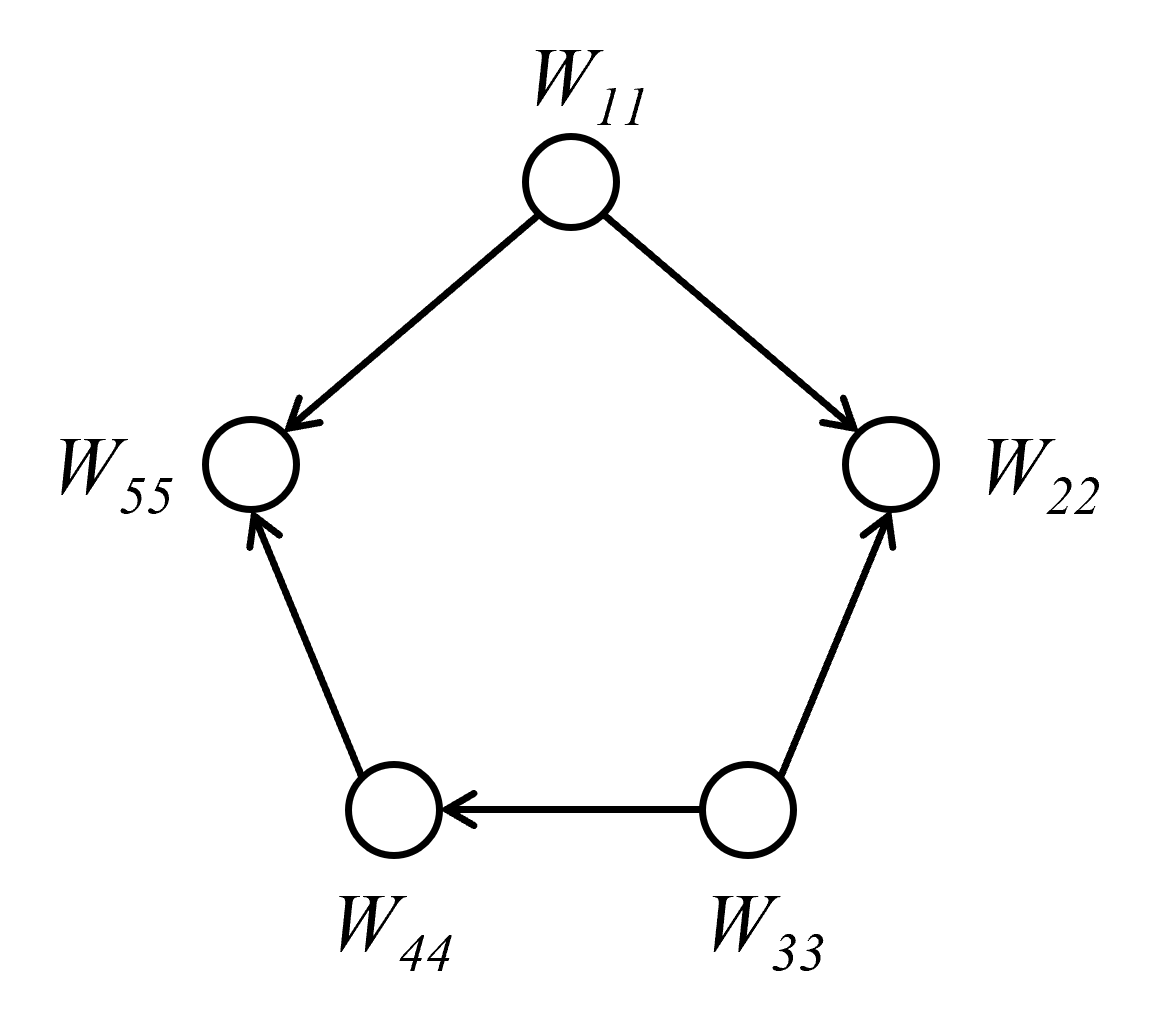}}  
   \hfil
   \subfloat[A one-to-one vector IA solution.]{%
     \label{5nodes_sample_conf_11vector_sol}
     \includegraphics[width=4.1cm]{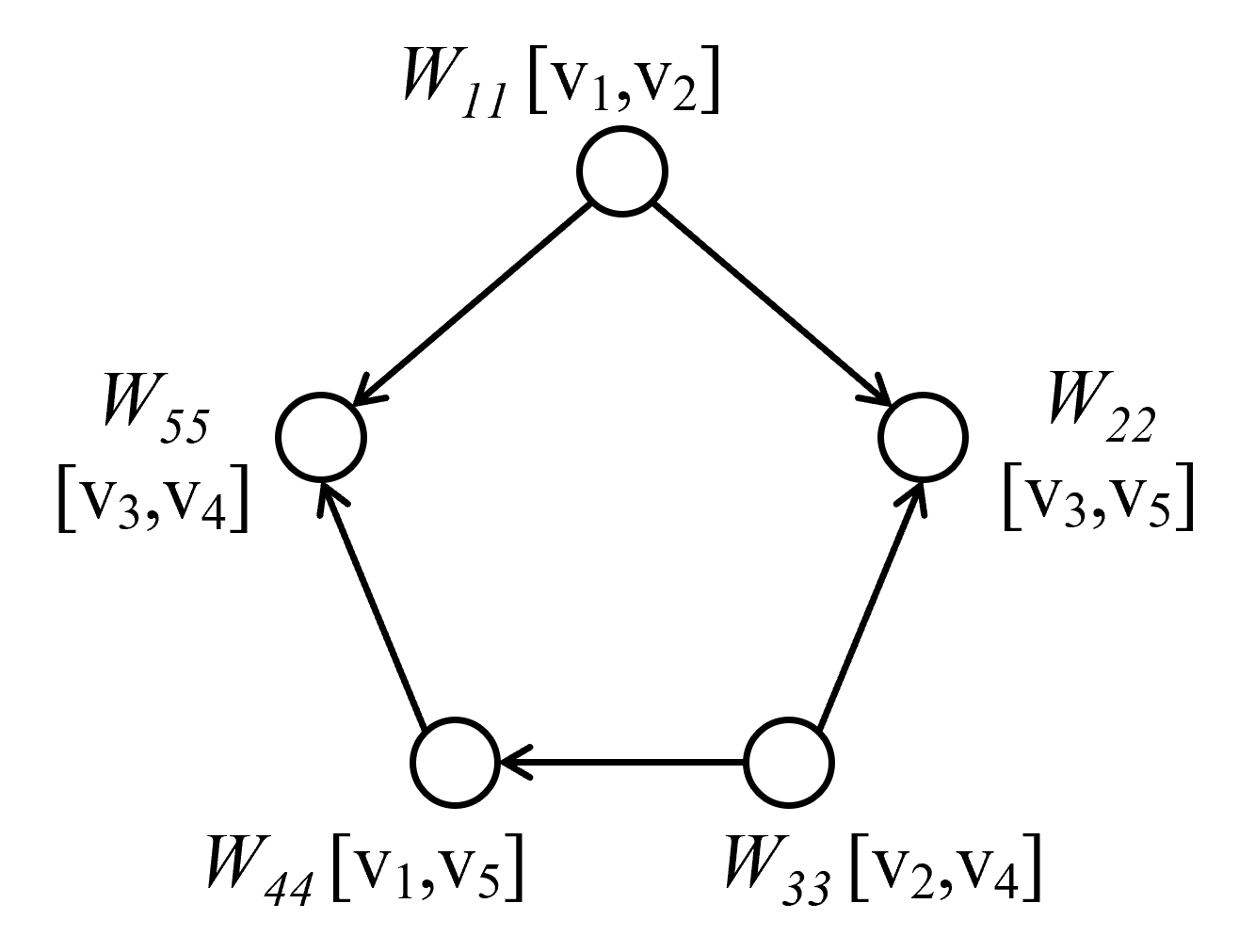}}  
   \caption{(a) A 5-node TIM instance topology graph, (b) the corresponding message conflict graph, and (c) a one-to-one vector IA solution.}
   \label{5_nodes_sample_vectorIA}
\end{figure*}

Among many linear coding schemes for TIM, interference alignment (IA) is one of the most promising approaches. IA aims to align signals from interfering transmitters as much as possible, while maintaining the signal from the desired transmitter separated. 
Based on how interfering messages are aligned, IA can be categorized into two classes: one-to-one IA and subspace IA. 
One-to-one IA is such that interferences are perfectly aligned in a one-to-one manner, where the beamforming vectors of aligned interfering symbols are identical.
 
Subspace IA, on the other hand, is to align interferences in a subspace with reduced dimensions.
In this case, the interfering symbols are not required to perfectly align one another on a one-to-one basis, but
one interfering symbol can align itself in the subspace spanned jointly by the beamforming vectors of other interfering symbols.

By combining the two classification criteria, we obtain four types of IA coding schemes: one-to-one scalar IA, one-to-one vector IA, subspace scalar IA, and subspace vector IA.

\subsubsection{One-to-One Scalar Interference Alignment (OSIA)}
Each source transmits a single symbol\footnote{The terminologies of symbol and message are used interchangeably for ease of presentation when referring to the scalar case.} with a corresponding beamforming vector in such a way that each interfering signal is perfectly aligned with another interfering one in the receiver's signal space, leaving one-dimensional interference-free subspace for each desired signal. 
From the receiver's viewpoint, each interfering transmitter sends a signal that is specifically designed to align with one dimension of the receiver's signal space, while the desired transmitter sends a signal that is independent of all the interfering signals. By aligning interfering signals within a reduced subspace of the receiver's signal space, OSIA can improve the achievable DoF over orthogonal access.

For instance, in Figure \ref{5nodes_sample_topo}, suppose each source sends one symbol via a beamforming vector. At $D_4$, the interference from $S_1$ and $S_3$ can be aligned by employing the same beamforming vector, which is linearly independent of that of the desired signal from $S_4$.

In the message conflict graph, OSIA can be interpreted as follows. The message $W_{44}$ sees two incoming edges from $W_{11}$ and $W_{33}$. If both messages $W_{11}$ and $W_{33}$ live in a subspace that does not contain $W_{44}$, then the desired message is separable from the interfering ones.

Therefore, the coding scheme design for scalar IA can be conducted by first generating a $K$-dimensional subspace spanned by linearly independent vectors $\{\v_k\}_{k=1}^K$, and then assigning one vector from this subspace to each node (i.e., message) in the message conflict graph.
To meet OSIA, the vector assignment should meet the following two conditions
\begin{itemize}
  \item [\textbf{C1})] Connected messages (nodes) should be assigned linearly independent vectors, regardless of the direction of edges.
  \item [\textbf{C2})] Messages (nodes) pointing to the same message (node) should be assigned as few different vectors as possible.
\end{itemize}

In the example of Figure \ref{5nodes_sample_conf}, we can generate a 2-dim subspace spanned by $\v_1=[1 \; 0]^T$ and $\v_2=[0 \; 1]^T$. Then, the vector assignment to the messages $W_{11}$, $W_{22}$, $W_{33}$, $W_{44}$, and $W_{55}$ will be $\v_1$, $\v_2$, $\v_1$, $\v_1+\v_2$, and $\v_2$, respectively. Note here that,
any pair of connected nodes, e.g. $W_{44}$ and $W_{55}$, are assigned linearly independent vectors, and the messages (e.g., $W_{11}$ and $W_{33}$) pointing to the same message $W_{44}$ are assigned the same vector for alignment. 

From the perspective of IA, the TIM coding turns out to be a vector assignment problem on message conflict graphs with a minimal vector subspace, whose dimensionality corresponds to the inverse of symmetric DoF, i.e., $\frac{1}{d_{\mathrm{sym}}}$.

\subsubsection{One-to-One Vector Interference Alignment (OVIA)}
Each source is allowed to transmit multiple symbols via several beamforming vectors, whereas the interfering signals are still aligned in a one-to-one manner. 
Different from the scalar case where the signals from different transmitters are either perfectly aligned or linearly independent, the vector case allows for partial alignment of interfering signals from different transmitters.
This enhances the flexibility of the design of interference alignment, leading to increased symmetric DoF compared to OSIA in some scenarios \cite{roth2006introduction}. 

As an example, Figure \ref{5nodes_sample_topo_11vector} depicts a TIM instance, where the use of OSIA results in a symmetric DoF of 1/3, which is equivalent to what can be achieved with orthogonal access such as time-division multiple access (TDMA). However, using the OVIA scheme, as demonstrated subsequently, we can increase the achievable symmetric DoF to 2/5. 

In the message conflict graph, the message $W_{55}$ sees two incoming edges from $W_{11}$ and $W_{44}$. The message $W_{22}$ sees two incoming edges from $W_{11}$ and $W_{33}$. If messages $W_{11}$, $W_{44}$ live in the subspace that does not contain $W_{55}$, and messages $W_{11}$, $W_{33}$ live in the subspace that does not contain $W_{22}$ then the desired message $W_{55}$ and $W_{22}$ could be separable from the interfering ones.

\begin{figure*}
   \centering
   \vspace{-20pt}
   \subfloat[Network topology graph.]{%
     \label{4nodes_sample_topo_subspacescalar}
     \includegraphics[width=3.1cm]{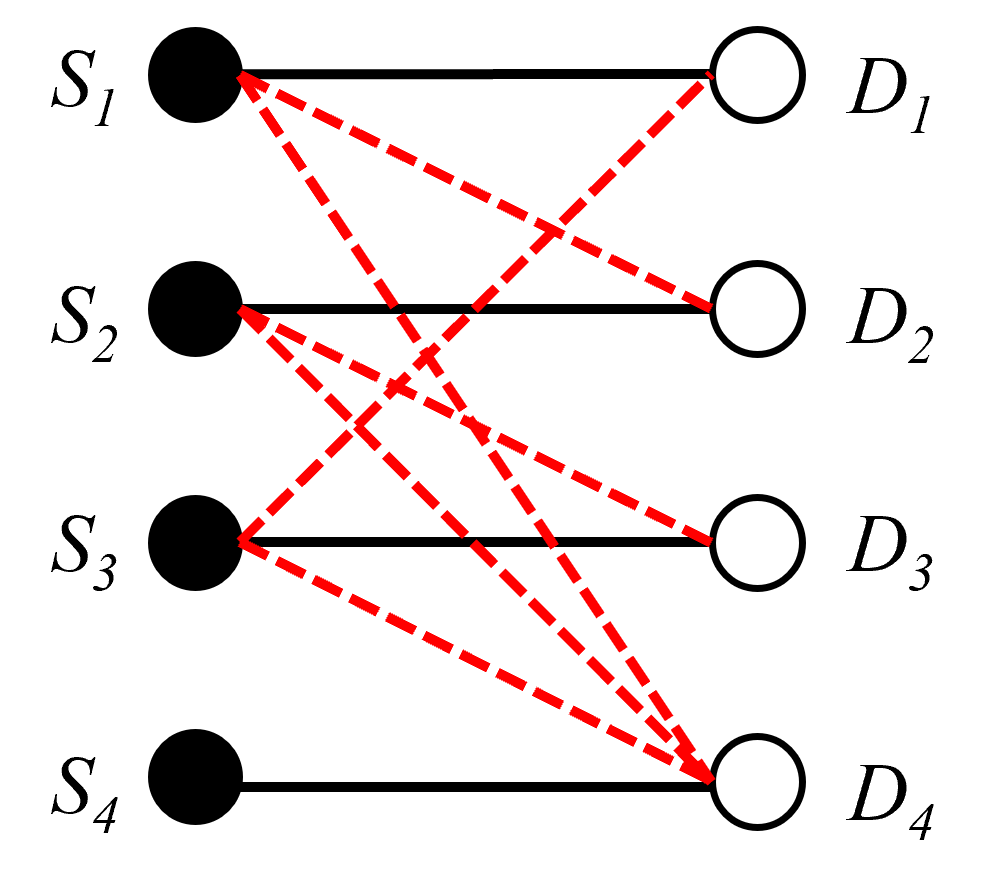}}
   \hfil
   \subfloat[Message conflict graph.]{%
     \label{4nodes_sample_conf_subspacescalar}
     \includegraphics[width=3.5cm]{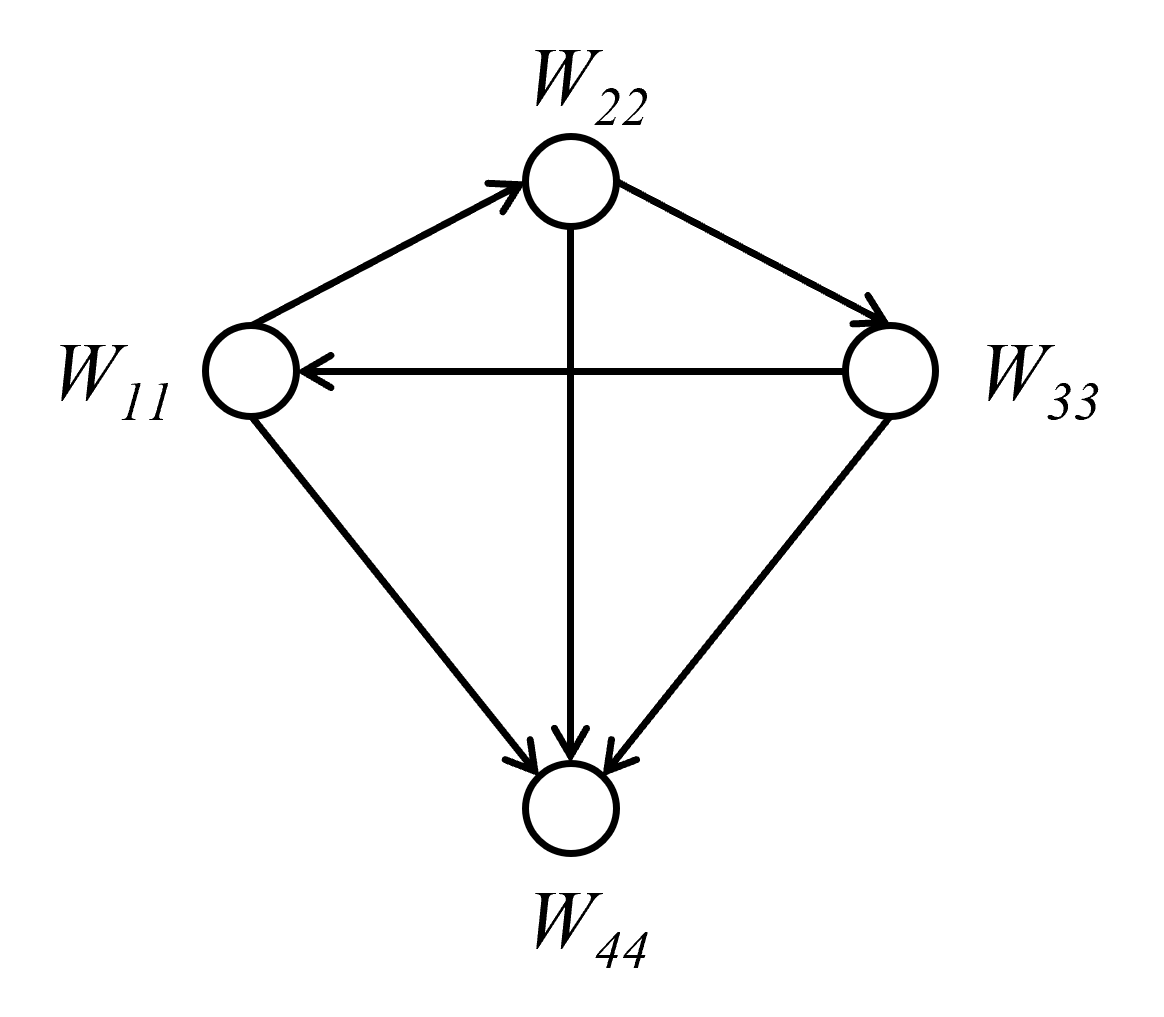}}  
   \hfil
   \subfloat[A subspace scalar IA solution.]{%
     \label{4nodes_sample_conf_subspacescalar_sol}
     \includegraphics[width=3.9cm]{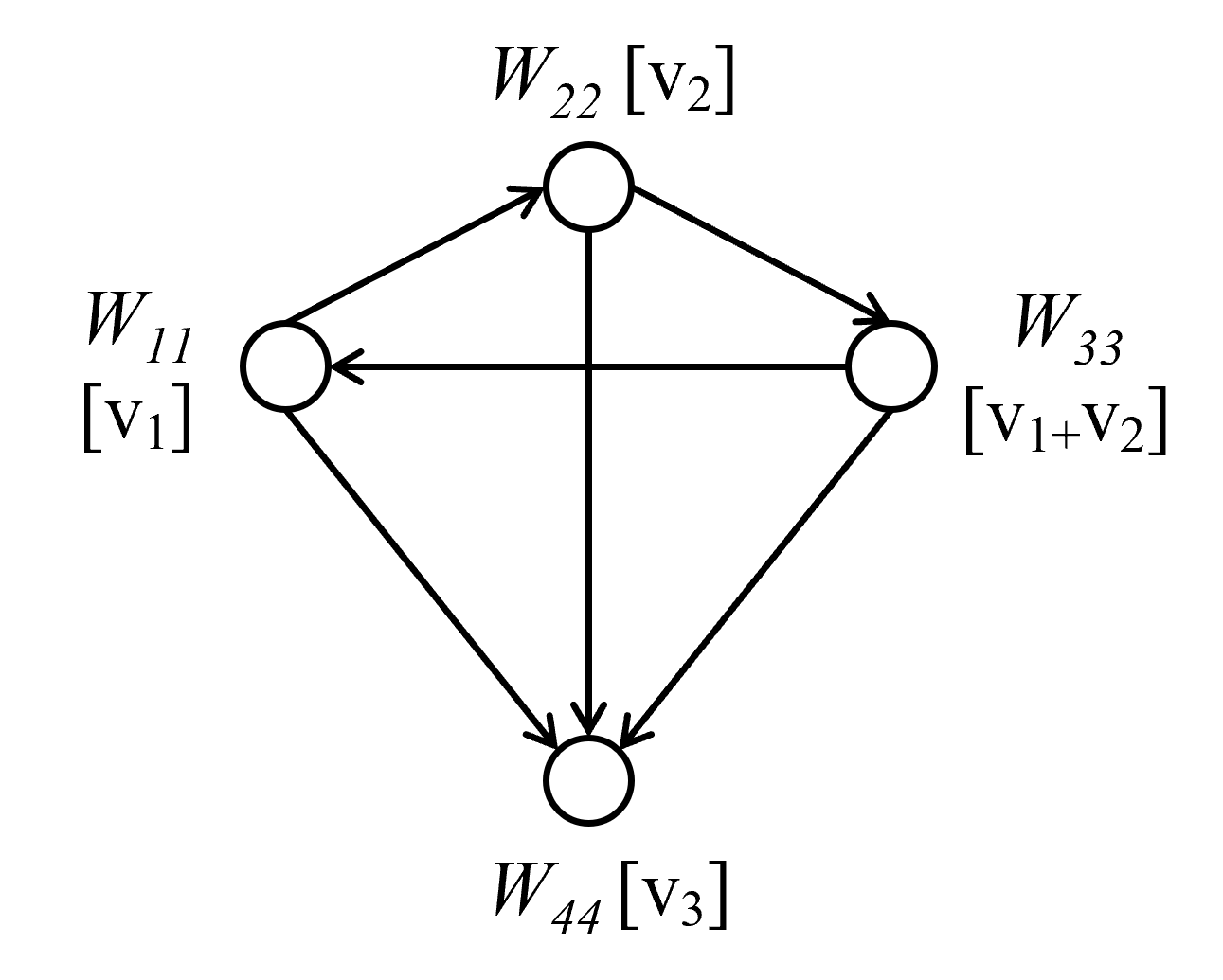}}  
   \caption{(a) A 4-node TIM instance topology graph, (b) the corresponding message conflict graph, and (c) a subspace scalar IA solution.}
   \label{4_nodes_sample_subspace_scalar}
\end{figure*}

Hence, the coding scheme design for vector IA involves generating a $K$-dimensional subspace spanned by linearly independent vectors $\{\v_k\}_{k=1}^K$ and assigning these vectors to nodes (i.e., messages) in the message conflict graph. It is worth noting that each node can be assigned multiple vectors. The vector assignment should follow the same conditions, \textbf{C1} and \textbf{C2}, as scalar IA.
Specifically, the OVIA coding scheme can be obtained as follows.
By randomly generating 5 vectors $\{\v_1,\v_2,\dots,\v_5\}$, each of size $5 \times 1$, that are in general linearly independent, over a sufficiently large field, we obtain the solution shown in Figure \ref{5nodes_sample_conf_11vector_sol} for the example depicted in Figure \ref{5nodes_sample_topo_11vector}.
Each message is split into two symbols, each of which is transmitted by some beamforming vector $\v$.
As a result, at the receiver side, the interfering signals occupy at most 3-dimensional subspace, e.g., 
$\dim([\v_1,\v_2],[\v_1,\v_5])=\dim([\v_1,\v_2],[\v_1,\v_4])=3$,
effectively avoiding all interference
and leaving at least a 2-dimensional interference-free subspace for the desired signal. This yields $d_{\mathrm{sym}}=\frac{2}{5}$ achievable.

\subsubsection{Subspace Scalar Interference Alignment (SSIA)}
Going beyond one-to-one IA, SSIA aims to align the interference from multiple interfering transmitters to a subspace of the receiver's signal space, while preserving a one-dimensional interference-free subspace that is separated from
the interference subspace, for the desired signal. In the SSIA, the signals from different interfering transmitters are not required to perfectly align one another, but rather interfering signals are restricted within a subspace with reduced dimensions. That is, one interfering signal may not perfectly align with another one, but falls in a subspace spanned by some other interfering signals.
This allows the receiver to isolate the interfering signals in that subspace and remove them from the desired signal. Meanwhile, the desired transmitter sends a signal that is independent of the interference subspace, occupying a different interference-free subspace. 

In certain scenarios, SSIA can increase the symmetric DoF of TIM over OSIA. 
By aligning the interfering signals to a subspace of the receiver's signal space, SSIA can effectively suppress interference from multiple sources, allowing for the simultaneous transmission of more independent data streams. 

Figure \ref{4_nodes_sample_subspace_scalar} presents an illustrative example in the context of TIM problems, where both OSIA and OVIA fail to achieve the symmetric DoF beyond $\frac{1}{4}$, while SSIA demonstrates the capability
to achieve the symmetric DoF of $\frac{1}{3}$. 
Specifically, the messages $W_{11}$, $W_{22}$, and $W_{33}$ all point to (interfere with) $W_{44}$, therefore they need to be aligned as much as possible. At the same time, these three messages form a directed cycle, indicating that they also need to be distinguished from each other pairwise. In this case, one of the messages, for example $W_{33}$, can be aligned with the subspace spanned by $W_{11}$ and $W_{22}$, while ensuring that they are linearly independent from each other. Meanwhile, the subspace spanned by $W_{11}$, $W_{22}$, and $W_{33}$ must do not contain $W_{44}$, which allows for the separation of all desired messages from the interfering ones.

If we define the set of all nodes pointing to a node $i$ in the graph as the in-neighborhood of node $i$, denoted as $N^+(i)$, then the vector assignment should follow the following two conditions to meet subspace IA:
\begin{itemize}
    \item [\textbf{C3})] Vectors assigned to each message (node) should not belong to the subspace spanned by the vectors assigned to its in-neighborhood.
    \item [\textbf{C4})]Vectors assigned to in-neighborhood of each node should occupy as small dimensional subspace as possible. 
\end{itemize}

In the example of Figure \ref{4nodes_sample_conf_subspacescalar_sol}, we can generate a 3-dim subspace spanned by three linearly independent vectors $\v_1=[1 \; 0\; 0]^T$, $\v_2=[0 \; 1\; 0]^T$ and $\v_3=[0 \; 0 \; 1]^T$. Then, we assign beamforming vectors $\v_1$, $\v_2$, $\v_1+\v_2$, and $\v_3$ to the messages $W_{11}$, $W_{22}$, $W_{33}$, and $W_{44}$, respectively. Note here that, the dimension of the interference subspace spanned by the vectors assigned to the in-neighborhood of $W_{44}$ is 2, i.e., $\dim(\operatorname{span}(\v_1, \v_2, \v_1+\v_2) = 2$, leaving 1-dim interference-free subspace to $W_{44}$. Meanwhile, $W_{11}$, $W_{22}$, $W_{33}$ are pairwise separable, as any two of their beamforming vectors $\v_1$, $\v_2$, and $\v_1+\v_2$ are linearly independent. This yields an achievable symmtric DoF of $\frac{1}{3}$, which is also optimal.

\subsubsection{Subspace Vector Interference Alignment (SVIA)}
When multiple symbols are transmitted from each source with a precoding matrix, SVIA 
seeks to align the interference from multiple interfering transmitters to a specific subspace within the receiver's signal space, while ensuring that a subspace orthogonal or independent of the interference subspace is preserved for the desired signals.

Specifically, let $b$ symbols be sent from each source with a $K \times b$ precoding matrix. Similarly to SSIA, SVIA does not require $b$ $K \times 1$ column vectors of the precoding matrix to be perfectly aligned with those from other sources in a one-to-one manner, but rather the subspace spanned by the column vectors is considered such that the interfering signals are forced to live in a subspace with reduced dimensionality.
As the only difference between SSIA and SVIA is the number of symbols each user could send, SVIA can be simply implemented by repeating SSIA multiple times with carefully designed beamforming vectors.

\section{Interference Alignment via Vector Assignment}
From the aforementioned IA coding schemes, it appears various IA coding can be implemented by properly generating beamforming vectors and assigning them to different messages with certain conditions \textbf{C1}-\textbf{C4} satisfied. In what follows, we consider the vector assignment strategies for both one-to-one and subspace IA.

\subsection{One-to-One IA via Local Coloring}
The main challenges of translating OSIA and OVIA into vector assignments consist of (1) the generation of linearly independent vectors that span a minimal vector subspace, and (2) the assignment of these vectors and their combinations to nodes in the message conflict graph.
If we interpret linearly independent vectors as different colors, then vector assignment can be alternatively done by color assignment, where the latter can be efficiently done via off-the-shelf graph coloring algorithms in the literature, e.g., \cite{matula1983smallest,deo2006interchange,hertz1987using}. 

In particular, we employ the techniques of graph local coloring and graph fractional local coloring to realize vector assignment for OSIA and OVIA, respectively.
\subsubsection{Vector Assignment via Local Coloring}\label{Local Coloring}

Vertex coloring aims to color the vertices of an undirected graph $\gG$ such that no two connected vertices are assigned the same color. This agrees with \textbf{C1} of IA.
The \textit{chromatic number} $\chi(\gG)$ of $\gG$ is the smallest number of colors needed.

Local vertex coloring is a special variant of vertex coloring on directed graphs, which was first introduced in \cite{KORNER2005101}.
It aims to color a directed graph $\gG_d = (\gV,\gE)$ with a vertex set $\gV$ and an edge set $\gE$, while minimizing the number of colors used in each local in-neighborhood. This agrees with \textbf{C2} of IA. The \textit{local chromatic number} $\chi_L(\gG_d)$ of $\gG_d$ is the smallest number of colors that appeared in the closed in-neighborhood of any vertex, over all valid vertex coloring on the underlying undirected graph.
The closed in-neighborhood $N_c^{+}(i)$ of a vertex $i$ is the union of the vertex $i$ and its in-neighborhood, i.e., $j\in N_c^{+}(i)$ iff $j = i$ or $(j,i) \in \gE$. Let $[m]$ denote the set \{$1,2,\dots,m$\} for some integer $m$. 

\begin{defn}[($K,r$)-local colorable]
Let $c:\gV \rightarrow [K]$ be any valid vertex coloring for the graph $\gG_d$ ignoring the direction of edges. Then, $\gG_d$ is $(K,r)$-local colorable if there exists some $c$ such that $\left|c(N_c^{+}(i)) \right| \leq r$ holds for all $i \in \gV$.
\end{defn}

When $K=\chi(\gG)$ and $r=\chi_L(\gG_d)$, we have optimal local coloring built upon optimal vertex coloring. Given such a coloring assignment, we can generate $K$ vectors with size $r$ each such that any $r$ of them are linearly independent. By assigning these vectors to all vertices of the message conflict graph, we end up with a valid OSIA coding scheme with $d_{\mathrm{sym}}=\frac{1}{\chi_L(\gG_d)}$, which is no less than $\frac{1}{\chi(\gG)}$ achieved by orthogonal access (e.g., TDMA) \cite{yi2018tdma} because $\chi_L(\gG_d) \le \chi(\gG)$.
 
\subsubsection{Vector Assignment via Fractional Local Coloring}\label{Fractional Local Coloring}
In fractional coloring, each vertex is assigned a set of colors rather than just one color, such that no two connected vertices share any common colors. A fractional coloring with set size $b$ is referred to as a $b$-fold coloring. An $a:b$-coloring refers to a $b$-fold coloring with $a$ available colors. The $b$-fold chromatic number, denoted as $\chi_b(\gG)$, is the smallest value of $a$ such that there exists an $a:b$-coloring. The fractional chromatic number, denoted as $\chi_f(\gG)$, is defined as follows:
\begin{align}
\chi_f(\gG) = \lim_{b \to \infty}\frac{\chi_{b}(\gG)}{b} = \inf_{b}\frac{\chi_{b}(\gG)}{b}.
\end{align}
When $b = 1$, fractional coloring reduces to a normal coloring.

Fractional local coloring is a fractional version of local coloring, aiming to minimize the number of colors used in each in-neighborhood for any valid fractional coloring. If we regard the $b$ colors assigned to a node as the $b$ beamforming vectors assigned to the corresponding message, then the requirements of fractional local coloring perfectly match those of OVIA. 
\begin{defn}[($K,r,b$)-fractional local colorable]
Let $c:\gV \rightarrow \{\{i_1,i_2,\ldots,i_b\} \mid 1 \leq i_1 < i_2 < \cdots < i_b \leq K\}$ be any valid fractional vertex coloring of size $b$ for the graph $\gG_d$ ignoring the direction of edges. Then, $\gG_d$ is $(K,r,b)$-fractional local colorable if there exists some $c$ such that $\left|c(N^{+}(i)) \right| \leq r$ holds for all $i \in \gV$.
\end{defn}

We further define the fractional local chromatic number $\chi_{fL}(\gG_d)$ of $\gG_d$ to be the smallest number of colors that appeared in the closed in-neighborhood of any vertex divided by $b$, over all valid fractional coloring on the underlying undirected graph i.e., $\min_c \frac{r}{b}$.

Given such a fractional color assignment, we can generate $K$ vectors with size $r \times 1$ each such that any $r$ of them are linearly independent. By assigning these vector sets to all vertices of the message conflict graph, we end up with a valid OVIA coding scheme with $d_{\mathrm{sym}}=\frac{1}{\chi_{Lf}(\gG_d)}$, which is no less than $\frac{1}{\chi_{L}(\gG)}$ achieved by OSIA, because $\chi_{Lf}(\gG_d) \le \chi_{L}(\gG)$ \cite{simonyi2006local}.

\subsubsection{Vector Generation via MDS Coding}
To generate $K$ vectors with size $r$ so that any $r$ of them are linearly independent, we resort to maximum distance separable (MDS) coding.

For a $q$-ary\footnote{The field size can be relaxed to infinite (i.e., the field of real/complex numbers) when considering the performance metric of DoF.} code with length $n$, width $k$, and minimum distance $d$, in short, $(n, k, d)_q$ code, the Singleton bound \cite{singleton1964maximum} states that $d \leq n-k+1.$ A code which meets this bound, i.e., $d = n-k+1$ is called MDS code, such as the Reed Solomon code. It provides the best error-correction capabilities due to their maximum distance among all codes. A $(n,k)$ MDS code is a set of $n$ vectors with length $k$. Every $k$ columns of the generator matrix $\mG_{k\times n}$ are linearly independent. 

\begin{figure}[htbp]
	\begin{center}
		\includegraphics[width=0.5\textwidth]{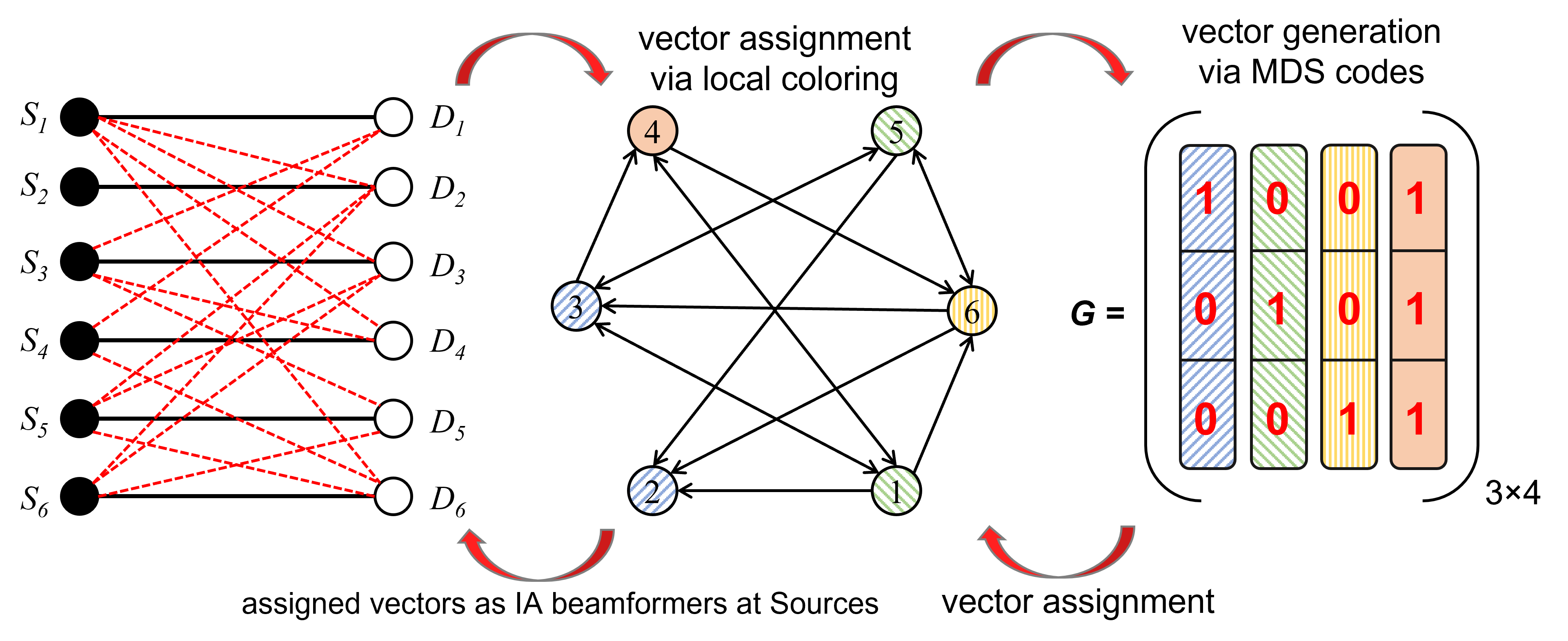}
	\end{center}
	\caption{An example of designing MDS code through local coloring to solve TIM instance. }
	\label{MDS_exaple}
\end{figure}

To summarize, OSIA via local coloring for TIM coding can be conducted as follows.
Given a TIM instance with message conflict graph $\gG_d$, and suppose a $(K,r)$-local coloring strategy on $\gG_d$ is already obtained, one can first create a $(K,r)$ MDS code with generator matrix $\mG_{r\times K}$. Then, by assigning each column of $\mG$ to each color class, the column assigned to any node $i$ is \textit{distinct} and \textit{linearly independent} from the columns assigned to the nodes pointed to $i$. The former comes from the rule of vertex coloring, and the latter is true because the number of colors in each in-neighborhood is at most $r$, and every $r$ columns of $\mG$ are linearly independent. Therefore, all destinations will be able to decode the demanded messages, as the interfering signals are perfectly aligned with a $(r-1)$-dim subspace, leaving 1-dim clean subspace to recover the desired message. 
An example of this process is shown in Figure \ref{MDS_exaple} with $K=4$ and $r=3$, and the IA coding yields $d_{\mathrm{sym}}=\frac{1}{r} = \frac{1}{3}$.

The process of conducting OVIA via fractional local coloring for TIM coding is analogous to the scalar case. Given a TIM instance with a message conflict graph $\gG_d$, and assuming a $(K, r, b)$-fractional local coloring strategy on $\gG_d$ has been obtained, one can first create a $(K, r)$ MDS code with a generator matrix $\mG_{r\times K}$. Subsequently, by assigning each column of $\mG$ to each color class, the $b$ columns assigned to any node $i$ will be distinct and linearly independent from the columns assigned to the nodes pointed to by $i$. This property is guaranteed by the rule of fractional vertex coloring, which ensures that adjacent vertices receive disjoint sets of columns, and the fact that every $r$ columns of $\mG$ are linearly independent. Consequently, all destinations will be able to decode the demanded messages, as the interfering signals are perfectly aligned within an $(r-b)$-dimensional subspace, leaving a $b$-dimensional clean subspace to recover the desired message. Hence, $d_{\mathrm{sym}}=\frac{b}{r}$. Note that when $b = 1$, OVIA reduces to OSIA.
Figure \ref{5_nodes_sample_vectorIA} is an example with $K = r = 5,b = 2$ and $d_{\mathrm{sym}}=\frac{2}{5}$.

\subsection{Subspace IA via Matrix Rank Reduction}\label{Subspace Interference Alignment via Matrix Rank Reduction}
While
subspace IA can still be implemented using vector assignment as
one-to-one IA, there are two different points: (1) Vector assignment should be done in a different way, as the graph coloring is not adequate to specify the overlap of different subspaces (cf. partial subspace alignment); (2) Vector generation should be done differently, as linearly independent vectors are not sufficient to meet the demands of subspace alignment, and therefore, MDS codes will not be used to generate beamforming vectors. 
To tackle these points, we propose a new vector assignment method with matrix rank reduction and a new way to generate beamforming vectors, which are tailored to subspace IA.

\subsubsection{Vector Assignment via Matrix Rank Reduction}
Let us assume that each node $i\in \gV$ is assigned $b$ beamforming vectors with size $x \times 1$, which are concatenated to form a matrix $A_i$, regardless of the order of vectors. Then we define the dimension of the subspace spanned by the assigned beamforming vectors in $i$'s in-neighborhood $N^+(i)$ as 
\begin{align}
r_N(i) = \mathrm{rank}( \oplus_{j \in N^+(i)} A_j ),
\end{align}
where $\oplus$ represents the matrix concatenation along the first dimension.
Similarly, we can define the dimension of the subspace spanned by the assigned beamforming vectors in $i$'s closed in-neighborhood $N_c^+(i)$ as 
\begin{align}
r_{cN}(i) = \mathrm{rank}(  \oplus_{j \in N_c^+(i)} A_j  ).
\end{align}

Next, we specify the requirements of matrix rank reduction as follows: 
\begin{itemize}
    \item [1)] For each node $i$ in $\G_d$, it must satisfy 
\begin{align}\label{condition1}
r_{cN}(i)-r_N(i)=b,
\end{align}
where $b$ is the number of assigned beamforming vectors for each node. This requirement meets the condition \textbf{C3}.
    \item [2)] $r_{cN}(i)$ to be as small as possible, meeting condition \textbf{C4}.
\end{itemize} 

We then define the so-called $(r,b)$-matrix rank reducible:
\begin{defn}[$(r,b)$-matrix rank reducible]
Let $c:\gV \rightarrow \vec{V}^b$ be any vertex assignment for the graph $\gG_d$, where $\vec{V}$ denotes the set of all vectors assigned to $\gG_d$. Then, $\gG_d$ is $r$-matrix rank reducible if there exists some $c$ such that both \eqref{condition1} and
\begin{align}
\max_{i \in \gV} r_{cN}(i) \leq r
\end{align}
hold.
\end{defn}

In fact, we can prove that a vector assignment scheme satisfying  (\ref{condition1}) is decodable. Since $\operatorname{span}(\oplus_{j \in N_c^+(i)} A_j) \supseteq \operatorname{span}(\oplus_{j \in N^+(i)} A_j)$, we have $r_{cN}(i)-r_N(i)=b$ if and only if $\mathrm{rank}(A_i) = b$ and $\operatorname{span}(A_i) \cap \operatorname{span}(\oplus_{j \in N^+(i)} A_j) = \emptyset$, for every $i$. Therefore, if (\ref{condition1}) is satisfied, the vectors assigned to each node $i$  are mutually independent and not contained in the subspace spanned by the assigned vectors in $N^+(i)$. Hence, each user $i$ would be able to decode the desired message.

We explain why we need to minimize $r_{cN}(i)$ to increase the symmetric DoF. Since $d_{\mathrm{sym}} = \frac{b}{x}$, to increase DoF as much as possible, we need to minimize the vector size $x$. For a given vector assignment scheme, we have 
\begin{align}
x \geq \max_{i \in \gV} r_{cN}(i).
\end{align}
Otherwise, there would not exist a matrix of any $x\times 1$ vectors with rank $r_{cN}(i)$. Therefore, we choose $x = \max_{i \in \gV} r_{cN}(i)$ to ensure that the vectors are separable, and hence, it is crucial to minimize $r_{cN}(i)$.

\subsubsection{Vector Generation}\label{vector_generation}
As mentioned earlier, vectors used for subspace IA usually require special design \cite{maleki2014index,suh2008interference}. Thanks to the efficiency of reinforcement learning, we do not need to design vectors in a special way as in traditional methods. Instead, we directly generate all ``0-1''  vectors of a given size $x \times 1$ corresponding to different subspace,
i.e., a total of $2^x - 1$ vectors excluding the all ``0'' vector. In the example of Figure \ref{4_nodes_sample_subspace_scalar}, we generate $2^3-1 = 7$ vectors, which correspond to all possible subspace
\begin{align}
\begin{pmatrix}
1 \\ 0 \\ 0
\end{pmatrix},
\begin{pmatrix}
0 \\ 1 \\ 0
\end{pmatrix},
\begin{pmatrix}
0 \\ 0 \\ 1
\end{pmatrix},
\begin{pmatrix}
1 \\ 1 \\ 0
\end{pmatrix},
\begin{pmatrix}
1 \\ 0 \\ 1
\end{pmatrix},
\begin{pmatrix}
0 \\ 1 \\ 1
\end{pmatrix},
\begin{pmatrix}
1 \\ 1 \\ 1
\end{pmatrix}.
\end{align}
It is worth noting that we use the placeholder $``1"$ to indicate that 
the corresponding subspace is occupied
and $``0"$ otherwise.  The task of assigning which vectors which node is delegated to the Matrix Rank Reduction algorithm. In practice, only a small subset of these vectors may be actually used.

\section{Learning to Code on Graphs}

In what follows, we propose different ways for the learning to code on graphs (LCG) framework to implement vector assignment for one-to-one and subspace IA.

\subsection{Learning for Local Coloring}
For the one-to-one IA,
while vector generation can be readily done by MDS code construction, it is challenging to assign vectors via local coloring in a systematic way. Even worse, there do not exist general local coloring algorithms in the literature. To address this challenge, we adopt a deep reinforcement learning (RL) approach \cite{ahn2020learning} to the $(K,r)$-local coloring problem.

\subsubsection{Overview of Our LCG Approach}\label{Transition}
The RL approach to the $(K,r)$-local coloring takes the directed graph $\gG_d$, message conflict graph, as the input and output a valid color assignment with in total $K$ colors and at most $r$ colors in the closed in-neighborhood of any node. 

Let $\gG_d=(\gV, \gE)$. 
First, we utilize a \textit{K-selector} to roughly estimate
$K$ as an input to RL, as shown in Figure \ref{1st framework}. For instance, one can use the greedy method to initialize $K$, and then repeat the RL method with decreasing $K$. At each iteration, the agent (policy network) gives a color from the color set to some of the undetermined vertices and defers the remaining vertices to later iterations. This process will be repeated until all vertices have been given colors.
The RL procedure can be represented as Markov decision processes (MDP) with four essential components:

\paragraph{State}
For each stage of the MDP, we define the RL state as a vertex-state vector $\vs = [s_i:i \in \gV]\in \{1,2,\dots,K,0\}^{\gV}$, where $s_i = k$ represents that vertex $i$ is assigned color ``$k$'', $k \in [K]$, and $s_i = 0$ represents vertex $i$ was decided to be deferred. The vertex-state is initialized to all ``deferred'', i.e., $\vs_0 = [0:i \in \gV]$, and the algorithm is terminated when all nodes have been colored or have reached the time limit $B$.

\paragraph{Action}
Given a state $\vs$, the agent will output a corresponding action $\va_0 = [a_i:i \in \gV_0]\in \{1,2,\dots,K,0\}^{\gV_0}$ towards deferred nodes set $\gV_0$. The nodes that have been colored in previous stages will not be given any new action. Similarly to the state, the vertex $i$ is assigned color ``$k$'' when $a_i = k$, or deferred when $a_i = 0$.

\paragraph{Transition}
There are two versions of transition: coloring and $(K,r)$-local coloring.

\begin{figure}[htbp]
	\begin{center}
		\includegraphics[width=0.45\textwidth]{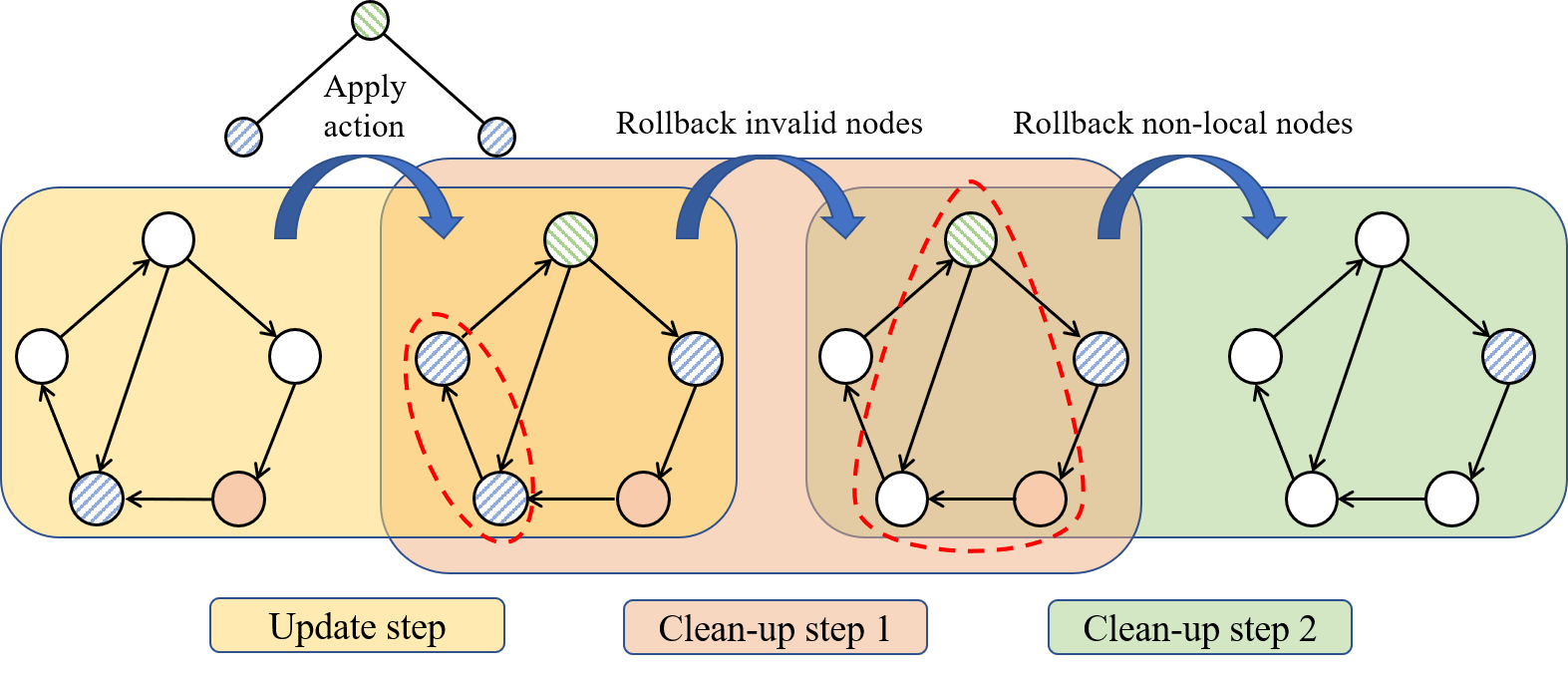}
	\end{center}
	\caption{Illustration of the transition process.}
	\label{transition2}
\end{figure}

\begin{itemize}
\item For coloring, the RL transits from state $\vs$ to the next state $\vs'$ through two steps: \textit{update} and \cleanone. In the \textit{update} step, RL overwrites the deferred part of the previous state with the action $\va_0$, resulting in an intermediate state $\vs''$, i.e., $s''_i = a_i$ for $i \in \gV_0$ and $s''_i = s_i$ otherwise. In the \cleanone\ step, RL identifies nodes that are adjacent but assigned the same color, which violates the rules of coloring. These nodes are then mapped back to ``deferred''. Note that even if the vertex $i$ is already colored in previous iterations, it remains possible to be rolled back.
\item For local coloring, the algorithm will perform one more step: \cleantwo. In this step, the closed in-neighborhood $N^+(i)$ of vertex $i$ whose number of in-color larger than $r$, i.e., $\{N^{+}(i): \left|c(N^{+}(i)) \right| > r , i \in \gV \}$, will be rolled back to the ``deferred''. Since such a $(K,r)$-local coloring scheme does not always exist, if the algorithm does not terminate after reaching the time limit $\alpha T$, clean-up-II will be discarded. This ensures that the model first tries to find a $(K,r)$-local coloring scheme, and if that fails, finds a $K$-coloring scheme. See Figure \ref{transition2} for a more detailed illustration of the transition between two states.
\end{itemize}

\paragraph{Reward}
We define the reward for taking an action in a state. The reward is the sum of two parts: the cardinality reward $R_{c}$, and the early-terminated reward $R_{t}$, i.e., 
\begin{IEEEeqnarray}{rCl}
R = R_{c}+\beta R_{t},
\end{IEEEeqnarray}
as described in Section \ref{reward}.

The policy network learns through repeated episodes (sequences of states, actions, and rewards) to adopt actions that maximize cumulative reward. Given the cumulative reward for each placement, we utilize proximal policy optimization (PPO) \cite{ppo} to update the policy network's parameters.

\subsubsection{Reward}\label{reward}
We consider two types of rewards.
\paragraph{Cardinality Reward}
Suppose the MDP transits from state $\vs$ to state $\vs'$, the cardinality reward is defined as 
\begin{IEEEeqnarray}{rCl}
R_{c}(\vs,\vs') = \sum_{i\in \gV\setminus \gV'_0} 1 - \sum_{i\in \gV \setminus \gV_0} 1.
\end{IEEEeqnarray}
This will reward the agent if more nodes are assigned in the new state. If an action causes more rollbacks than are assigned, the model gets a negative reward. By doing so, RL tends to extend the cardinality of the successfully assigned node set.

\paragraph{Early-terminated Reward}
In order to encourage the model to make decisions as fast as possible, the model is rewarded with $R_{t}$ when the algorithm terminates at time $t$ and given the time limit $B$, where 
\begin{IEEEeqnarray}{rCl}
R_{t} = \frac{B-t}{B}.
\end{IEEEeqnarray}
Our experiments have demonstrated that this significantly increases the speed of training.

\subsubsection{Policy and Value Network Architecture}
Our model uses Actor-Critic reinforcement learning based on graph convolutional neural networks (GCNN). Both policy network $\pi(\va|\vs)$ and value network $q(\vs,\va)$ consist of $4$-layers GraphSAGE networks \cite{graphsage} with GCN aggregator \cite{gcn}. The $n$-th layer performs
the following transformation on input $\mH$: 
\begin{IEEEeqnarray}{rCl}
h^{(n)}(\mH) = \text{ReLU}(\mH \mW_1^{(n)}+\mD^{-\frac{1}{2}} \mB \mD^{-\frac{1}{2}}\mH \mW_2^{(n)}),
\end{IEEEeqnarray}
where $\mB$ and $\mD$ represent the adjacency matrix and degree matrix, respectively. $\mW_1^{(n)}$ and $\mW_2^{(n)}$ are the weights updated during the training process. To create actions and value estimations at the final layer, the policy and value networks use softmax and graph read-out functions with sum pooling \cite{gin} instead of ReLU. The neural network's input features are the current iteration-index of the MDP and the sum of the one-hot encoding of the neighbor's state. Thanks to these features, we only take the subgraph induced on the deferred vertices $\gV_0$ as the input of the networks.

\subsection{Learning for Fractional Local Coloring}
For one-to-one vector IA,
we utilize the technique of node splitting to transform 
fractional local coloring in the original graph into a more conventional local coloring problem on a reconstructed graph. Through the construction of a refined $b$-order node splitting graph, we embark on a local coloring endeavor within this intricately crafted graph. Each node undergoes meticulous splitting, assigning a solitary color to each constituent node within the splitting graph. However, the key of this approach lies in the amalgamation of the split nodes belonging to each original node, thereby unveiling a remarkable fractional local coloring scheme for the original graph. This transformation grants us the privilege of seamlessly applying the well-established LCG method, which we previously proposed, to efficiently resolve the local coloring predicament within the node splitting graph, ultimately surmounting the challenges posed by the fractional local coloring problem in the original graph.

\begin{defn}[$b$-order node splitting graph]
Let $\gG = (\gV, \gE)$ be a directed graph, and let $b$ be a positive integer. The $b$-order node splitting graph of  $\gG$, denoted as $\gG_b' = (\gV_b', \gE_b')$, is defined as follows:
Each original node $v \in \gV$ is split into $b$ nodes $v_1, v_2, \ldots, v_b$ such that 
\begin{align}
    \gV_b' = \{v_1, v_2, \ldots, v_b \mid v \in \gV \}.
\end{align}
For each edge $(u, v) \in \gE$, there exist $b^2$ edges $\{(u_i, v_j) \mid 1 \leq i,j \leq b \}$ in $\gE_b'$. Moreover, the $b$ split nodes of each original node are fully connected. That is, 
\begin{align}
    \gE_b' = &\{(u_i, v_j) \mid (u, v) \in \gE, 1 \leq i,j \leq b \} \\
    &\cup \{(v_i, v_j) \mid v \in \gV, 1 \leq i,j \leq b, i \neq j\}.
\end{align}
\end{defn}

In other words, the b-order node splitting graph\footnote{The b-order node splitting graph can be seen as the Cartesian product of the original conflict graph and the directed clique of size $b$.} is obtained by splitting each node in the original graph into $b$ directed fully connected nodes, and then connecting each split node of $u$ to each split node of $v$, if $u$ points to $v$ in the original graph. 

\begin{figure}[htbp]
	\begin{center}
		\includegraphics[width=0.48\textwidth]{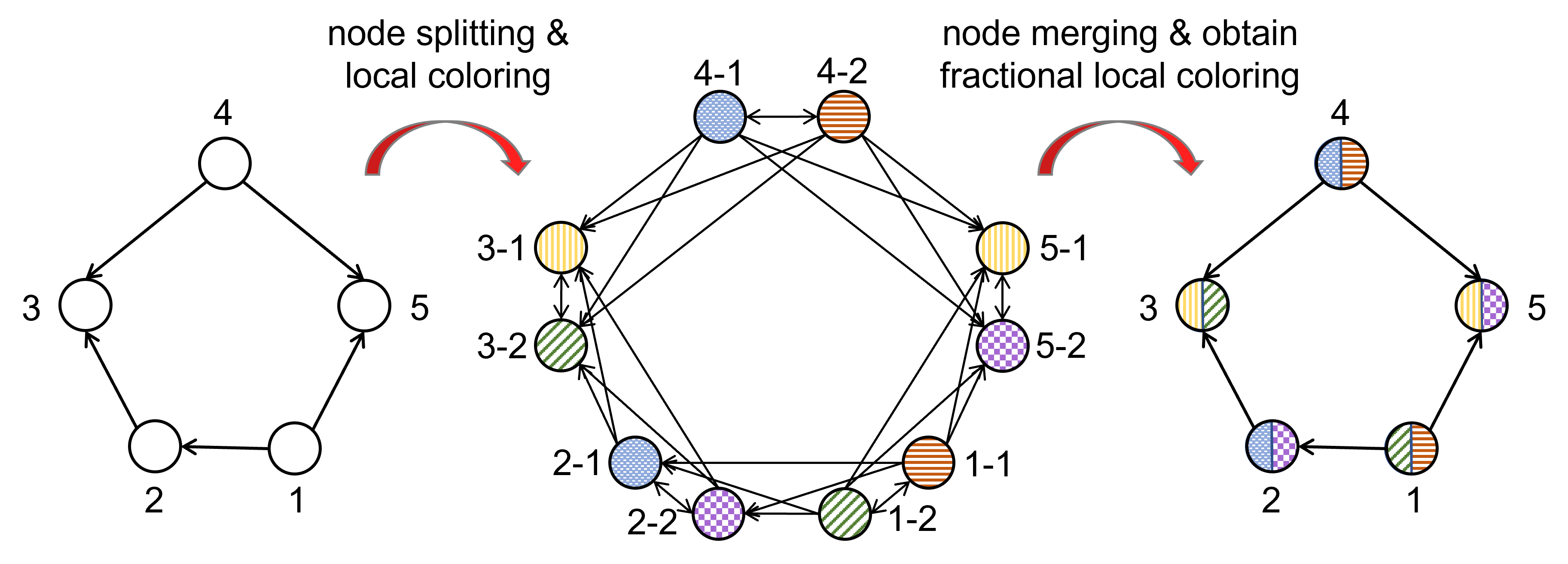}
	\end{center}
	\caption{Illustration of node splitting and merging.}
	\label{node_splitting_picture}
\end{figure}

Figure \ref{node_splitting_picture} showcases an example for the generation of a $2$-order node splitting graph, the subsequent application of local coloring on the splitting graph, and the subsequent merger of the splitting graph and coloring scheme back into the original graph to yield a fractional local coloring scheme. 

The proof establishing that a valid ($K,r$)-local coloring of a $b$-order node splitting graph $\gG_b'$ can be seamlessly merged to produce a valid ($K,r,b$)-fractional local coloring of the original graph $\gG$ is straightforward and can be derived directly from the underlying definitions. In essence, we have effectively expanded the scope of the LCG method to address the fractional local coloring problem by employing a combination of node splitting and merging techniques.

\subsection{Learning for Matrix Rank Reduction}\label{Learning_for_Matrix_Rank_Reduction}
For the subspace IA,
our proposed LCG framework to address the $(r,b)$-matrix rank reduction problem follows a similar one used for solving the local coloring problem. However, there are specific differences in the construction of the MDP to accommodate the matrix rank reduction condition. These differences primarily manifest in the design of new state-decision pairs and the transition mechanism tailored for matrix rank reduction. It is important to note that in our analysis, we assume $b=1$, corresponding to the subspace scalar IA, as the case of $b>1$ can be effectively resolved using the node splitting technique, similar to resolving fractional local coloring.

In the RL approach to matrix rank reduction, the message conflict graph $\gG_d=(\gV, \gE)$ and the rank parameter $r$ are provided as inputs. The RL procedure outputs a vector assignment scheme satisfying the eq. \ref{condition1} with a vector size of $x=r \times 1$. To start the process, we generate a vector set $\vec{V}$ that consists of all possible vectors. The size of this vector set is $2^r - 1$, as explained in Section \ref{vector_generation}. At each iteration, the agent selects some vectors from $\vec{V}$ and assigns them to some of the undetermined vertices (one vertix will be assigned at most one vector) while deferring the remaining vertices for subsequent iterations. This iterative process continues until all vertices have been assigned vectors.

The components of MDP are following:

\paragraph{State}
For each stage of the MDP, we define the RL state as a vertex-state vector $\vs = [s_i:i \in \gV]\in \{1,2,\dots,2^r-1,0\}^{\gV}$, where $s_i = k$ represents that vertex $i$ is assigned vector $\v_k$, $\v_k \in \vec{V}$, and $s_i = 0$ represents vertex $i$ was decided to be deferred. The vertex-state is initialized to all ``deferred'', i.e., $\vs_0 = [0:i \in \gV]$, and the algorithm is terminated when all nodes have been assigned or have reached the time limit $B$.

\paragraph{Action}
Given a state $\vs$, the agent will output a corresponding action $\va_0 = [a_i:i \in \gV_0]\in \{1,2,\dots,2^r-1,0\}^{\gV_0}$ towards deferred nodes set $\gV_0$. The nodes that have been assigned vector in previous stages will not be given any new action. The vertex $i$ is assigned vector $\v_k$ when $a_i = k$, or deferred when $a_i = 0$.

\paragraph{Transition}
The RL transits from state $\vs$ to the next state $\vs'$ through two steps: \textit{update} and \cleanup. The \textit{update} step is the same as the one in \ref{Transition}. In the \cleanup\ step, RL computes the values of $r_N(i)$ and $r_{cN}(i)$ for each vertex $i$, and then rolls back the state of the vertex and its in-neighborhood to ``deferred'' if the condition \ref{condition1}, $r_{cN}(i) - r_N(i) = 1$, is not satisfied.

The reward function and network architecture remain unchanged, and therefore will not be further elaborated upon in this context.

\section{Experiments}
In this section, the proposed LCG framework is evaluated with extensive experiments for coloring, local coloring, fractional local coloring, and matrix rand reduction against a variety of TIM instances.\footnote{The detailed experiment setups and 
 implementation are available at \url{https://github.com/ZhiweiShan/Learning-to-Code-on-Graphs}}
 
\textbf{Dataset generation}:
To thoroughly assess the efficacy of the proposed LCG method in comparison to other approximation or heuristic algorithms for graph coloring schemes, we generate diverse types of random bipartite graphs as network topology graphs. Specifically, we consider Erdős-Rényi (ER) graphs \cite{batagelj2005efficient}, preferential attachment (PA) graphs \cite{barabasi1999emergence}, Havel-Hakimi (HH) graphs \cite{havel1955remark,hakimi1962realizability}, and wireless network simulation graphs (Wireless Net) \cite{yi2015itlinq+}. Additionally, we directly generate random geometric (GEO) graphs \cite{penrose2003random} and Barabasi-Albert (BA) graphs \cite{albert2002statistical} as message conflict graphs to assess their transferability.

To classify the datasets, we determine their chromatic numbers by solving linear programming problems using Gurobi \cite{gurobi}. The size of each bipartite graph is presented in pairs, representing the number of nodes in both node sets. For instance, $(30, 30)$ denotes a network comprising 30 source nodes and 30 destination nodes.

For a comprehensive understanding of the graph instance generation process, we refer the reader to Appendix \ref{Dataset_Details}, where complete details are provided.

\textbf{LCG with RL}:
 In our LCG approach, we generate a total of 50,000 graphs for training and 5,000 graphs for evaluation, employing various random parameter settings. These graphs are subsequently categorized into multiple datasets based on their chromatic numbers, which serve as the basis for both the training and evaluation stages. The hyperparameters utilized in our approach can be found in Appendix \ref{Implementation_of_L2C}. Notably, the number of colors $K$ is set equal to the chromatic number of each dataset.

It is important to mention that, for testing all the local coloring, fractional local coloring and matrix rank reduction, we employ the model trained on the coloring problem using ER graphs. The only distinction lies in utilizing the corresponding versions of the Markov Decision Process (MDP) transition, as outlined in Section \ref{Transition} and \ref{Learning_for_Matrix_Rank_Reduction}.

\textbf{Baselines}:
We compare our method to two heuristic algorithms: the smallest-last greedy method with the interchange (SLI) \cite{matula1983smallest, deo2006interchange} and TabuCol \cite{hertz1987using}. The smallest-last greedy method is a simple yet powerful approach. It assigns each vertex in a sequential manner to the lowest indexed color that does not result in any conflicts, and it adds new colors when necessary. The interchange technique is employed to enhance the effectiveness of any sequential coloring algorithm.

TabuCol, on the other hand, is a well-studied heuristic based on Tabu local search. In this method, the chromatic number is assumed to be given, and we set the maximum number of iterations for TabuCol to 1000.

All our experiments are performed using a single GPU (NVIDIA A100 40 GB) and a single CPU (AMD EPYC 7452).
\subsection{Experiments for Coloring}
To assess the quality of the solutions, we measure the optimal ratio, which represents the proportion of graphs that can be colored using $\chi(\gG)$ colors out of the total number of graphs. The performance results are presented in Table \ref{performance}, encompassing various general coloring problems.

Notably, our LCG model consistently achieves the highest optimal ratios across all datasets, outperforming other methods. Moreover, LCG exhibits superior computational efficiency compared to the SLI greedy method in most cases. Although the SLI algorithm demonstrates relatively poorer performance, it offers the advantage of faster execution time, as anticipated, when compared to TabuCol.

\begin{table}[htbp]
	\centering
	\caption{Optimal ratio on test graphs, where the best ratios are marked in bold. Running times (in seconds) are provided in brackets. In this table, training and testing use data from the same distribution. }
	\label{performance}
	\begin{footnotesize}		
		\begin{tabular}{ccc|ccc}
			\hline
			Type                          & N                      & $\chi$ & SLI         & TabuCol     & LCG         \\ \hline
			\multirow{4}{*}{ER}           & \multirow{2}{*}{(15, 15)} & 5      & 0.98 (1.83) & \textbf{1} (9.58)    & \textbf{1} (2.38)    \\
			&                        & 6      & 0.99 (0.68) & \textbf{1} (2.53)    & \textbf{1} (1.79)    \\ \cline{2-6} 
			& \multirow{2}{*}{(30, 30)} & 7      & 0.73 (7.28) & 0.87 (1677) & \textbf{0.92} (5.10) \\
			&                        & 8      & 0.85 (15.8) & 0.92 (2699) & \textbf{0.94} (8.59) \\ \hline
			\multirow{4}{*}{PA}           & \multirow{2}{*}{(15, 15)} & 5      & \textbf{1} (4.58)    & \textbf{1} (25.9)    & \textbf{1} (2.55)    \\
			&                        & 6      & \textbf{1} (4.38)    & \textbf{1} (23.9)    & \textbf{1} (2.53)    \\ \cline{2-6} 
			& \multirow{2}{*}{(30, 30)} & 7      & 0.99 (4.46) & \textbf{1} (100)     & \textbf{1} (2.98)    \\
			&                        & 8      & \textbf{1} (5.41)    & \textbf{1} (103)     & \textbf{1} (3.47)    \\ \hline
			\multirow{4}{*}{HH}           & \multirow{2}{*}{(15, 15)} & 5      & \textbf{1} (5.23)    & \textbf{1} (44.8)    & \textbf{1} (3.05)    \\
			&                        & 6      & \textbf{1} (4.54)    & \textbf{1} (36.8)    & \textbf{1} (2.85)    \\ \cline{2-6} 
			& \multirow{2}{*}{(30, 30)} & 7      & 0.99 (5.35) & \textbf{1} (258)     & \textbf{1} (3.58)    \\
			&                        & 8      & 0.99 (9.19) & \textbf{1} (356)     & \textbf{1} (5.22)    \\ \hline
			\multirow{4}{*}{Wireless Net} & \multirow{2}{*}{(15, 15)} & 5      & 0.94 (9.64) & \textbf{1} (406)     & \textbf{1} (5.06)    \\
			&                        & 6      & \textbf{1} (10.6)    & \textbf{1} (169)     & \textbf{1} (5.41)    \\ \cline{2-6} 
			& \multirow{2}{*}{(30, 30)} & 7      & 0.88 (28.5) & 0.97 (3602) & \textbf{0.99} (13.2) \\
			&                        & 8      & 0.99 (10.8) & 0.99 (497)  & \textbf{1} (6.36)    \\ \hline
		\end{tabular}
	\end{footnotesize}
\end{table}

\subsection{Experiments for Local Coloring}
We evaluate the performance of LCG in solving the $(K,K-1)$-local coloring problem. As LCG is the first algorithm designed for this specific problem, we do not include comparisons with other algorithms. We conduct local coloring tests on wireless networks with sizes of (15, 15) and (30, 30).

\begin{figure}[htbp]
   \centering
   \subfloat[(15, 15) nodes Wireless Net.]{
    \label{fig:subfigA}
     \includegraphics[width=4.3cm]{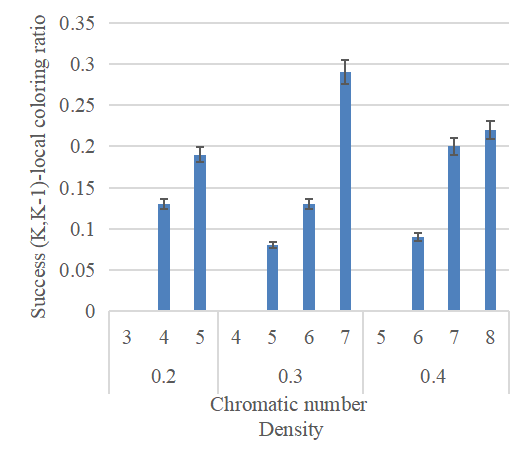}}
   \hfil
   \subfloat[(30, 30) nodes Wireless Net.]{
     \label{fig:subfigB}
     \includegraphics[width=4.3cm]{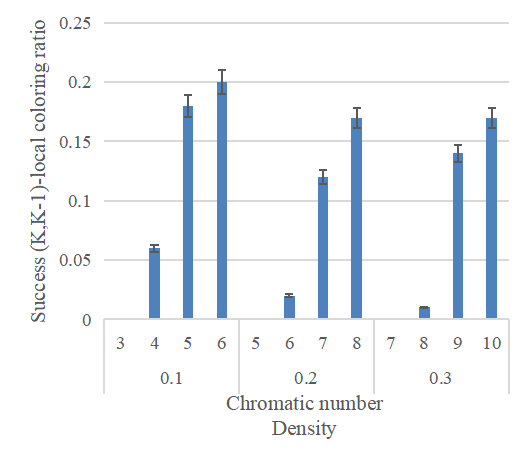}}  
   \caption{Successful local coloring ratio on Wireless Network graphs with different size and density. Each dataset contains 100 graphs.}
   \label{local Wireless Net}
\end{figure}

Figure~\ref{local Wireless Net} provides a summary of the results. Subfigure (a) shows the results for the (15, 15) nodes Wireless Net, while subfigure (b) shows the results for the (30, 30) nodes Wireless Net. Each dataset consists of 100 graphs.

For example, in the (15, 15) nodes dataset with a density of 0.3 and a chromatic number of 7, the highest achieved ratio is approximately 0.29. This indicates that approximately 29\% of the generated TIM instances exhibit suboptimal performance with TDMA and can benefit from using IA coding schemes instead of orthogonal access \cite{yi2018tdma}.
\subsection{Experiments for Fractional Local Coloring}
We test the ability of LCG with node splitting to handle the $(K,r,b)$-fractional coloring problem, i.e. OVIA. We perform the tests on a randomly generated set of 431 Wireless Nets with N = (8, 8) and $\chi = 3$, considering various values of $b$ and $r$. The original graph was directly used for testing OSIA, while it was transformed into a 2-order node splitting graph and a 3-order node splitting graph for testing 2-dim OVIA and 3-dim OVIA, respectively.

We present the number of cases achieving different $d_{\mathrm{sym}} = \frac{b}{r}$ values under different methods (i.e., different choices of $b$) in Table \ref{node_splitting}. From the results presented, we can see that LCG with node splitting is an effective approach for addressing the $(K,r,b)$-fractional coloring problem. Specifically, while both OSIA and 2-dim OVIA were able to achieve 4 cases with $d_{\mathrm{sym}} = 1/2$, 2-dim OVIA was able to achieve an additional $13-4 = 9$ cases with $d_{\mathrm{sym}} = 2/5$, making it a stronger method than OSIA, as expected. Additionally, 3-dim OVIA was able to achieve $5-4 = 1$ additional case with $d_{\mathrm{sym}} = 3/7$, making it stronger than both OSIA and 2-dim OVIA. However, it should be noted that those 13 cases with $d_{\mathrm{sym}} = 3/8$ achieved a higher $d_{\mathrm{sym}} = 2/5$ in 2-dim OVIA. This is because these methods are only able to produce fractional results, and not continuous ones. Based on this, we can speculate that the optimal $d_{\mathrm{sym}}$ for these examples lie in the range $[2/5, 3/7)$.
\begin{table}[htbp]
\centering
\caption{Number of Cases Achieving Different Symmetric DoF under OSIA and OVIA. ``/" indicates not applicable}
\label{node_splitting}
\resizebox{0.7\columnwidth}{!}{
\begin{tabular}{c|ccccc}
\hline
\diagbox{Method}{$d_{\mathrm{sym}}$}  & 1/3 & 3/8 & 2/5 & 3/7 & 1/2 \\ \hline
OSIA               & 431 & /   & /   & /   & 4   \\ \hline
2-dim OVIA         & 431 & /   & 13  & /   & 4   \\ \hline
3-dim OVIA         & 431 & 13  & /   & 5   & 4   \\ \hline
\end{tabular}
}
\end{table}

\subsection{Experiments for Matrix Rank Reduction}
We test the ability of LCG to handle the $(r,1)$-matrix rank reduction problem, i.e., SSIA. We perform the tests on a randomly generated set of 577 Wireless Nets with N = (15, 15) and $\chi = 4$, considering $r=4,3$. 
As a comparison, we also tested the performance of $(4,r)$-fractional coloring on the same dataset. As shown in the Table \ref{SSIA}, both methods were able to achieve $d_{\mathrm{sym}}=1/4$ for all cases, which is the limit that conventional orthogonal access approaches can achieve. For $d_{\mathrm{sym}}=1/3$, OSIA was able to achieve it for 10\% of the cases (60 out of 431), while SSIA was able to achieve it for 18\% of the cases (106 out of 431), which is 8\% more cases than OSIA.

\begin{table}[htbp]
\centering
\caption{Number of Cases Achieving Different Symmetric Dof under OSIA and SSIA}
\label{SSIA}
\resizebox{0.5\columnwidth}{!}{
\begin{tabular}{c|cc}
\hline
\multicolumn{1}{c|}{\diagbox{Method}{$d_{\mathrm{sym}}$}} & \multicolumn{1}{c}{1/4} & \multicolumn{1}{c}{1/3} \\ \hline
OSIA & 577 & 60 \\ \hline
SSIA & 577 & 106 \\ \hline
\end{tabular}
}
\end{table}

\subsection{Performance Evaluation for Wireless Networks}
Given the focus of the TIM problem on wireless networks, wireless network topologies hold particular significance. Leveraging the high generalization and transferability of our LCG model, we train it on ER graphs with (15, 15) and (30, 30) nodes and evaluate its performance on Wireless Net graphs with nodes ranging up to (100, 100). A comprehensive summary of the results is presented in Table \ref{Wireless}.

The outcomes clearly demonstrate the feasibility of training LCG on small-scale ER graphs and utilizing it effectively for large-scale network simulation graphs. This exemplifies the scalability and adaptability of our model across diverse wireless network scenarios. For more specific information regarding the parameters employed in generating wireless network topologies, please refer to Appendix \ref{Dataset_Details}.

\begin{table}[htbp]
	\centering
	\caption{Optimal ratio for wireless network simulation graphs. Each test dataset is filtered from 5000 test graphs according to $\chi$.}
	\label{Wireless}
	\begin{tabular}{cc|c|ccc}
		\hline
		\multicolumn{2}{c|}{\begin{tabular}[c]{@{}c@{}}Training on\\ ER graph\end{tabular}} & \multicolumn{4}{c}{Test on Wireless Net graph} \\ \hline
		N                      & $\chi$             & N       & SLI         & TabuCol      & LCG         \\ \hline
		\multirow{6}{*}{(15, 15)} & \multirow{3}{*}{5} &  (30, 30)   & 0.85 (20.8) & 0.98 (976)   & \textbf{1} (12.4)    \\
		&                    &  (50, 50)   & 0.96 (23.2) & 0.99 (10421) & \textbf{1} (20.7)    \\
		&                    & (100, 100) & 0.99 (39.5) & 0.99 (7142)  & \textbf{1} (29.5)    \\ \cline{2-6} 
		& \multirow{3}{*}{6} &  (30, 30)   & 0.99 (13.1) & \textbf{1} (263)      & \textbf{1} (10.1)    \\
		&                    &(50, 50)   & 0.99 (8.98) & \textbf{1} (1136)     & \textbf{1} (7.32)    \\
		&                    & (100, 100) & 0.99 (8.42) & \textbf{1} (593)      & \textbf{1} (6.54)    \\ \hline
		\multirow{6}{*}{(30, 30)} & \multirow{3}{*}{7} &  (30, 30)   & 0.88 (25.6) & 0.97 (1718)  & \textbf{0.99} (21.8) \\
		&                    & (50, 50)   & 0.97 (34.9) & \textbf{0.99} (1727)  & \textbf{0.99} (13.8) \\
		&                    & (100, 100) & 0.92 (12.3) & 0.92 (7729)  & \textbf{0.99} (11.7) \\ \cline{2-6} 
		& \multirow{3}{*}{8} &  (30, 30)   & 0.99 (15.4) & \textbf{1} (250)      & \textbf{1} (10.4)    \\
		&                    & (50, 50)   & \textbf{1} (6.41)    & \textbf{1} (160)      & \textbf{1} (3.83)    \\
		&                    & (100, 100) & 0.99 (3.22) & 0.99 (560)   & \textbf{1} (4.14)    \\ \hline
	\end{tabular}
\end{table}

\subsection{Generalization and Transferability}
Finally, we conduct an evaluation of the generalization and transferability of our method, specifically examining its performance on unseen graph types and varying sizes. To investigate this, we train the LCG model on ER graphs of different sizes and subsequently test its generalization ability on ER, HH, GEO, and BA graphs. The comprehensive results are presented in Table \ref{Generalization}.

The outcomes clearly demonstrate that our LCG model exhibits excellent performance on unseen graph types and sizes. This inherent capability is advantageous as it allows us to train LCG on a specific graph type without the need to consider the graph type during testing. This flexibility greatly enhances the applicability and versatility of our method across diverse graph scenarios.
\begin{table}[htbp]
	\centering
	\caption{Optimal ratio on graphs from different types and size graphs. Each dataset is filtered from 5000 test graphs by $\chi=7$.}
	\label{Generalization}
	\resizebox{\columnwidth}{!}{
	\begin{tabular}{l|ccccccc}
		\hline
		\diagbox{Train}{\\Test} &
		\begin{tabular}[c]{@{}c@{}}ER\\ (15, 15)\end{tabular} &
		\begin{tabular}[c]{@{}c@{}}ER\\ (20, 20)\end{tabular} &
		\begin{tabular}[c]{@{}c@{}}ER\\ (25, 25)\end{tabular} &
		\begin{tabular}[c]{@{}c@{}}ER\\ (30, 30)\end{tabular} &
		\begin{tabular}[c]{@{}c@{}}HH\\ 20\end{tabular} &
		\begin{tabular}[c]{@{}c@{}}Geo\\ 50\end{tabular} &
		\begin{tabular}[c]{@{}c@{}}BA\\ 50\end{tabular} \\ \hline
		ER (15, 15) & 0.999 & 0.995 & 0.996 & 0.934 & 1 & 1 & 0.998 \\ \hline
		ER (20, 20) & 1     & 0.993 & 0.996 & 0.944 & 1 & 1 & 0.996 \\ \hline
		ER (25, 25) & 0.997 & 0.993 & 0.996 & 0.928 & 1 & 1 & 0.997 \\ \hline
		ER (30, 30) & 1     & 0.993 & 0.995 & 0.925 & 1 & 1 & 0.995 \\ \hline
	\end{tabular}
}
\end{table}

\section{Conclusion}
\label{sec:conclusion}
Building upon the relation between interference alignment and local graph coloring, we proposed a learning to code on graphs (LCG) framework for the TIM problems, leveraging deep reinforcement learning for graph coloring. By exploiting local coloring of message conflict graphs, LCG automatically assign colors (coding vectors) to different messages so as to achieve one-to-one interference alignment (IA).
The proposed LCG framework was further extended to  the vector version of one-to-one IA and subspace IA to discover new advanced IA coding schemes.
A comprehensive experimental evaluation of the proposed framework demonstrates its effectiveness in coloring and local coloring.
As the first learning-to-code approach to the TIM problem, we hope this work could stimulate the future development of coding techniques, bringing in new advances from machine learning. 

\IEEEtriggeratref{100}

\bibliographystyle{IEEEtran}
\bibliography{paper} 

\clearpage
\appendices

\section{Dataset Details}\label{Dataset_Details}
We train on graphs randomly generated based on variant specific parameters. 50,000 graphs were generated and separated according to the chromatic number for training, and 5,000 graphs for evaluation.
\begin{itemize}
    \item  The specific parameters for random graphs in Table \ref{performance}, \ref{Wireless},
    \ref{Generalization} are shown in Table \ref{Table_1_parameter}, \ref{Table_2_parameter}, \ref{Table_3_parameter} separately. $q$ represents the percent of randomly choosing demanded messages.
    
    \item For Wireless Net, we randomly distribute transmitters and receivers within a square area of 1,000 m $\times$ 1,000 m.  As in \cite{yi2015itlinq+}, the simulated channel follows the LoS model in ITU-1411. The carrier frequency is 2.4 GHz, antenna height is 1.5 m, and the antenna gain per device is -2.5 dB. The noise power spectral density is -174 dBm/Hz, and the noise figure is 7 dB. Each pair of transmitter and receiver is uniformly spaced within [2, 65] meters. Each link is expected to operate over a 10 MHz spectrum and the maximum transmit power is 30 dBm. 
\end{itemize}

\begin{table}[htbp]
\caption{Specific parameters for generating graph in Table \ref{performance}. Wireless Net follows all-unicast setting, therefore q does not apply.}
\label{Table_1_parameter}
\resizebox{\columnwidth}{!}{
\begin{tabular}{cc|c|cll}
\hline
Type & N                         & q & \multicolumn{3}{c}{Specific parameters}                                           \\ \hline
\multirow{4}{*}{ER} &
  \multirow{2}{*}{(15, 15)} &
  \multirow{12}{*}{0.2} &
  \multicolumn{3}{c}{\multirow{2}{*}{Probability for edge creation = 0.2}} \\
     &                           &   & \multicolumn{3}{c}{}                                                              \\ \cline{2-2} \cline{4-6} 
     & \multirow{2}{*}{(30, 30)} &   & \multicolumn{3}{c}{\multirow{2}{*}{Probability for edge creation = 0.2}}          \\
     &                           &   & \multicolumn{3}{c}{}                                                              \\ \cline{1-2} \cline{4-6} 
\multirow{4}{*}{PA} &
  \multirow{2}{*}{(15, 15)} &
   &
  \multicolumn{3}{c}{\multirow{2}{*}{\begin{tabular}[c]{@{}c@{}}Probability that a new bottom node is added = 0.2\\ Max degree of the random degree sequence = 7\end{tabular}}} \\
     &                           &   & \multicolumn{3}{c}{}                                                              \\ \cline{2-2} \cline{4-6} 
 &
  \multirow{2}{*}{(30, 30)} &
   &
  \multicolumn{3}{c}{\multirow{2}{*}{\begin{tabular}[c]{@{}c@{}}Probability that a new bottom node is added = 0.2\\ Max degree of the random degree sequence = 6\end{tabular}}} \\
     &                           &   & \multicolumn{3}{c}{}                                                              \\ \cline{1-2} \cline{4-6} 
\multirow{4}{*}{HH} &
  \multirow{2}{*}{(15, 15)} &
   &
  \multicolumn{3}{c}{\multirow{2}{*}{Max degree of the random degree sequence = 6}} \\
     &                           &   & \multicolumn{3}{c}{}                                                              \\ \cline{2-2} \cline{4-6} 
     & \multirow{2}{*}{(30, 30)} &   & \multicolumn{3}{c}{\multirow{2}{*}{Max degree of the random degree sequence = 8}} \\
     &                           &   & \multicolumn{3}{c}{}                                                              \\ \hline
\multirow{4}{*}{\multirow{2}{*}{\begin{tabular}[c]{@{}c@{}}Wireless\\ Net\end{tabular}}} &
  \multirow{2}{*}{(15, 15)} &
  \multirow{4}{*}{\textbackslash{}} &
  \multicolumn{3}{c}{\multirow{2}{*}{\begin{tabular}[c]{@{}c@{}}Topological density = 0.4\end{tabular}}} \\
     &                           &   & \multicolumn{3}{c}{}                                                              \\ \cline{2-2} \cline{4-6} 
 &
  \multirow{2}{*}{(30, 30)} &
   &
  \multicolumn{3}{c}{\multirow{2}{*}{\begin{tabular}[c]{@{}c@{}} Topological density = 0.3\end{tabular}}} \\
     &                           &   & \multicolumn{3}{c}{}                                                              \\ \hline
\end{tabular}
}
\end{table}

\begin{table}[htbp]
\caption{Specific parameters for generating graph in Table \ref{Generalization}.}
\label{Table_2_parameter}
\resizebox{\columnwidth}{!}{
\begin{tabular}{cc|c|cll}
\hline
Type & N        & q & \multicolumn{3}{c}{Specific parameters}                          \\ \hline
\multirow{4}{*}{ER} &
  (15, 15) &
  \multirow{5}{*}{0.2} &
  \multicolumn{3}{c}{Probability for edge creation = 0.3} \\ \cline{2-2} \cline{4-6} 
     & (20, 20) &   & \multicolumn{3}{c}{Probability for edge creation = 0.25}         \\ \cline{2-2} \cline{4-6} 
     & (25, 25) &   & \multicolumn{3}{c}{Probability for edge creation = 0.2}          \\ \cline{2-2} \cline{4-6} 
     & (30, 30) &   & \multicolumn{3}{c}{Probability for edge creation = 0.2}          \\ \cline{1-2} \cline{4-6} 
HH   & (20, 20)  &   & \multicolumn{3}{c}{Max degree of the random degree sequence = 7} \\ \hline
GEO &
  50 &
  \multirow{2}{*}{\textbackslash{}} &
  \multicolumn{3}{c}{Distance threshold value = 0.2} \\ \cline{1-2} \cline{4-6} 
BA &
  50 &
   &
  \multicolumn{3}{c}{\begin{tabular}[c]{@{}c@{}}Number of edges to attach from \\ a new node to existing nodes = 7\end{tabular}} \\ \hline
\end{tabular}
}
\end{table}

\begin{table}[h!]
\caption{Specific parameters for generating graph in Table \ref{Wireless}.}
\label{Table_3_parameter}
\resizebox{\columnwidth}{!}{
\begin{tabular}{cc|c|cll}
\hline
Type &
  N &
  q &
  \multicolumn{3}{c}{Specific parameters} \\ \hline
\multirow{2}{*}{ER} &
  (15, 15) &
  \multirow{2}{*}{0.2} &
  \multicolumn{3}{c}{Probability for edge creation = 0.2} \\ \cline{2-2} \cline{4-6} 
 &
  (30, 30) &
   &
  \multicolumn{3}{c}{Probability for edge creation = 0.2} \\ \hline
\multirow{3}{*}{\multirow{2}{*}{\begin{tabular}[c]{@{}c@{}}Wireless\\ Net\end{tabular}}} &
  (30, 30) &
  \multirow{3}{*}{\textbackslash{}} &
  \multicolumn{3}{c}{\begin{tabular}[c]{@{}c@{}}Channel magnitude percentile threshold value = 0.2 (0.3)\\ for chromatic number 5, 6 (7, 8)\end{tabular}} \\ \cline{2-2} \cline{4-6} 
 &
  (50, 50) &
   &
  \multicolumn{3}{c}{\begin{tabular}[c]{@{}c@{}}Channel magnitude percentile threshold value = 0.1 (0.15)\\ for chromatic number 5, 6 (7, 8)\end{tabular}} \\ \cline{2-2} \cline{4-6} 
 &
  (100, 100) &
   &
  \multicolumn{3}{c}{\begin{tabular}[c]{@{}c@{}}Channel magnitude percentile threshold value = 0.04 (0.08)\\ for chromatic number 5, 6 (7, 8)\end{tabular}} \\ \hline
\end{tabular}
}
\end{table}

\section{Implementation of LCG}\label{Implementation_of_L2C}
We use the same hyperparameters for each experiments, except for the maximum iterations per episode $B$. Evaluation result in Table \ref{performance} and Table \ref{Generalization} is obtained by setting $B = 32$, while that of Table \ref{Wireless} is obtained by setting $B = 64$. Policy and value networks were parameterized using a graph convolutional network with four layers and 128 hidden dimensions. Using the Adam optimizer with a learning rate of 0.001, each instance of the model was trained for 5,000 iterations of proximal policy optimization \cite{ppo}. For each instance, we calculate 20 cases in parallel and select the best results to report. The gradient norms were clipped by $0.2$. As a function of the number of vertices in a dataset, the cardinality reward is normalized by this number.

\end{document}

%% file: math_commands.tex
\usepackage{amsmath,amsfonts,bm}
\def\cleanup{\textit{clean-up}}
\def\cleanone{\textit{clean-up-\uppercase\expandafter{\romannumeral1}}}
\def\cleantwo{\textit{clean-up-\uppercase\expandafter{\romannumeral2}}}

\def\eqref#1{equation~\ref{#1}}

\def\1{\bm{1}}

\def\va{{\bm{a}}}

\def\vs{{\bm{s}}}

\def\mB{{\bm{B}}}

\def\mD{{\bm{D}}}

\def\mG{{\bm{G}}}
\def\mH{{\bm{H}}}

\def\mT{{\bm{T}}}

\def\mW{{\bm{W}}}

\DeclareMathAlphabet{\mathsfit}{\encodingdefault}{\sfdefault}{m}{sl}
\SetMathAlphabet{\mathsfit}{bold}{\encodingdefault}{\sfdefault}{bx}{n}

\def\gE{{\mathcal{E}}}

\def\gG{{\mathcal{G}}}

\def\gM{{\mathcal{M}}}

\def\gV{{\mathcal{V}}}

